\RequirePackage[2020-02-02]{latexrelease}
\documentclass[aps,prd,superscriptaddress,showpacs,preprint,amsmath,amssymb]{revtex4}
\usepackage{graphicx, bm}
\usepackage{float}
\usepackage{subcaption}
\usepackage[usenames]{color}

\begin{document}

%\tightenlines
\draft

\title{The new $Z'$ boson of the Bestest Little Higgs Model as a portal to signatures of Higgs bosons $h_0$ and $H_0$ at the future muon collider }

\author{J. M. Mart\'inez-Mart\'inez\footnote{jose.martinez@fisica.uaz.edu.mx}}
\affiliation{\small Facultad de F\'{\i}sica, Universidad Aut\'onoma de Zacatecas\\
         Apartado Postal C-580, 98060 Zacatecas, M\'exico.\\}

\author{ A. Guti\'errez-Rodr\'{\i}guez\footnote{alexgu@fisica.uaz.edu.mx}}
\affiliation{\small Facultad de F\'{\i}sica, Universidad Aut\'onoma de Zacatecas\\
         Apartado Postal C-580, 98060 Zacatecas, M\'exico.\\}

\author{ E. Cruz-Albaro\footnote{elicruzalbaro88@gmail.com}}
\affiliation{\small Facultad de F\'{\i}sica, Universidad Aut\'onoma de Zacatecas\\
         Apartado Postal C-580, 98060 Zacatecas, M\'exico.\\}

\author{ M. A. Hern\'andez-Ru\'{\i}z\footnote{mahernan@uaz.edu.mx}}
\affiliation{\small Unidad Acad\'emica de Ciencias Qu\'{\i}micas, Universidad Aut\'onoma de Zacatecas\\
         Apartado Postal C-585, 98060 Zacatecas, M\'exico.\\}

\date{\today}
%\maketitle

\begin{abstract}
% Insert abstract here

We study the new $Z'$ boson as a portal for the production of Higgs bosons $h_0$ and $H_0$ predicted by the Bestest Little Higgs Model
through the Higgs-strahlung processes $\mu^+\mu^- \to (Z, Z')  \to Zh_0, ZH_0$. We focus on the resonance and non-resonance effects of
the $Zh_0, ZH_0$ signals. In our analysis, we consider the center-of-mass energies of $\sqrt{s}=3, 4, 6,  10, 30$ 
%$\sqrt{s}=3, 4, 5, 6, 7, 10, 30$
TeV and integrated luminosities of ${\cal L}=2, 4, 6, 10, 30$ ${\rm ab^{-1}}$ projected for a future muon collider. The possibility of performing precision measurements for the Higgs bosons $h_0$ and $H_0$ is very promising at the future muon collider.  Furthermore, our results may be helpful to the High Energy Physics community. Complementarily, we generate and provide the Feynman rules necessary for studying the processes $\mu^+\mu^- \to (Z, Z')  \to Zh_0, ZH_0$. 

\end{abstract}

%\pacs{12.60.-i, 14.70.Hp, 14.70.Bh \\

\vspace{5mm}

\maketitle

%\narrowtext

\section{Introduction}

Since the confirmation of the existence of the Higgs boson by the ATLAS \cite{ATLAS:2012yve} and the CMS \cite{CMS:2012qbp} Collaborations
at the Large Hadron Collider (LHC), the scientific community has undertaken the task of confirming that its properties match those of the scalar boson predicted by the Standard Model (SM)  \cite{ATLAS:2015zhl,CMS:2014nkk,ATLAS:2015egz,CMS:2014fzn,ATLAS:2016neq}.
Any deviation from the SM predictions could be valuable information about a possible extended theory.
 Despite the great predictive power of the SM, there are unsolved problems, such as the hierarchy problem. Namely, quantum corrections render the Higgs potential fine-tuned. These quantum corrections come from three different sectors of the SM: the gauge sector, the fermion sector, and the Higgs sector. To have a theory that is not fine-tuned, new physics is needed in each of these three sectors to cancel quantum corrections to the Higgs potential.

Several physics models beyond the SM have been proposed to solve the so named hierarchy problem. Some of the proposed extensions are the Little Higgs Models (LHM)~\cite{Arkani-Hamed:2002ikv,Chang:2003zn,Han:2003wu,Chang:2003un,Schmaltz:2004de} that employ a mechanism called collective symmetry breaking. Its main idea is to represent the SM Higgs boson as a pseudo-Nambu-Goldstone boson of an approximate global symmetry spontaneously broken at a certain scale in the teraelectronvolt (TeV) range. 
In these models, the collective symmetry-breaking mechanisms are implemented in three sectors: fermion, gauge, and Higgs.  In each sector, new particles in the mass range of a few TeV are predicted.  
 These new particles play the role of partners of the top quark, of the gauge bosons, and the Higgs boson, the effect of which is to generate radiative corrections for the mass of the Higgs boson and, thus cancel the divergent corrections induced by SM particles. 
 On the other hand, the LHM~\cite{Arkani-Hamed:2002ikv,Chang:2003zn,Han:2003wu,Chang:2003un,Schmaltz:2004de} have the inconvenience of being strongly constrained by electroweak precision measurements in the gauge sector~\cite{JHEP09-2010,Csaki:2002qg,Csaki:2003si}, but also predict  top partners that are much heavier than the mass of the new gauge bosons, which leads to significant fine-tuning in the Higgs potential~\cite{JHEP09-2010,Casas:2005ev}. 

The Bestest Little Higgs Model (BLHM)~\cite{JHEP09-2010,Kalyniak:2013eva,Godfrey:2012tf,Cisneros-Perez:2023foe,Cruz-Albaro:2023pah,Cruz-Albaro:2022lks,Cruz-Albaro:2022kty,Cruz-Albaro:2024vjk,Cisneros-Perez:2024efk,Cisneros-Perez:2024onx} overcomes the difficulties presented by LHM, which is achieved by incorporating two independent symmetry-breaking scales,  $ f $ and $ F $ with $ F>f$.
 Consequently, a disassociation in the masses of the quark ($ T $, $ T_5 $, $ T_6 $, $ T^{2/3} $, $ B $, $ T^{5/3} $) and boson gauge ($ Z' $, $ W'^{\pm} $) partners is generated.
  Regarding the quarks, four of them are heavy partners of the top quark ($+2/3$ charge), one heavy partner of the bottom quark ($-1/3$ charge), and one exotic quark  ($+5/3$ charge).
  The new quarks obtain masses proportional only to the $ f $ energy scale, while the new gauge bosons acquire masses proportional to the combination of the $ f $ and $ F $ scales, i.e. $\sqrt{f^{2}+F^{2}}$.
 Since the new quarks are now lighter than the new gauge bosons, fine-tuning in the fermion sector and electroweak precision data constraints in the gauge sector are avoided.
  On the other hand, the scalar sector of the BLHM has a rich phenomenology that generates the scalar fields: $h_0, H_0, A_0, \phi^{0},\eta^{0}, H^{\pm}, \phi ^{\pm}$ and $ \eta^{\pm}$. The $h_0$ state is assumed to be similar to the SM Higgs boson.
For more information on the BLHM, we recommend that the interested reader refer to Refs.~\cite{Cisneros-Perez:2023foe,Cruz-Albaro:2023pah,Cruz-Albaro:2022lks,Cruz-Albaro:2022kty,Cruz-Albaro:2024vjk,Cisneros-Perez:2024efk,Cisneros-Perez:2024onx}.

Given that the SM Higgs boson plays essential roles in several extended model scenarios, it is natural to expect that new physics beyond the SM would influence the properties of this Higgs boson,  thus leading to deviations of Higgs properties from SM predictions.   Probing the extended models meaningfully beyond the direct LHC searches would require precision measurements of the Higgs boson couplings. So far, several couplings of the Higgs to SM fermions and vector bosons based on current LHC data still have large uncertainties, for example, the $ZZh_0$ coupling~\cite{Dawson:2013bba,Han:2013kya,Bechtle:2014ewa}.
Such a precision, if achieved, will be very useful to discover the evidence of
new physics beyond the SM.
On the other hand, Lepton colliders have the advantage of  clean signatures and high-statistics samples of the Higgs boson.
 Compared to the Hadron collider, a future Lepton collider may have a higher capacity in the measurement of the $ZZh_0$ coupling through Higgs-strahlung production $l^{+}l^{-} \to Zh_0$.
 The  $l^{+}l^{-}$ colliders could reduce the aforementioned uncertainties to  few percent level.

In this paper, we explore the phenomenology of the production of the Higgs bosons $h_0$ and $H_0$ of the BLHM in muon collisions. Specifically, we will present a comprehensive analysis of the production mechanism $\mu^{+}\mu^{-}\rightarrow (Z, Z') \to Zh_0$ and $\mu^{+}\mu^{-}\rightarrow (Z, Z') \to ZH_0$ and its sensitivity including both the resonant and the nonresonant effects at future high-energy and high-luminosity muon collider.
In the BLHM scenario, the Higgs-strahlung productions $\mu^+ \mu^- \to Zh_0$ and $\mu^+ \mu^- \to  ZH_0$ are essential processes to study tree-level interactions: $Z Z h_0$, $Z Z H_0$, $Z'Z h_0$, and $Z'Z H_0$.  At the same time, the above processes are useful for testing the consistency of the parameter space of the BLHM.
 Our search for the Higgs bosons is implemented in the environment of a future muon collider~\cite{MuonCollider:2022nsa,Accettura:2023ked}, as this could provide a potential solution to the issues regarding energy, luminosity, background cleanliness, and the limited sensitivity of current and other future colliders.  
These features make the muon collider an ideal collider for the search for new particles, as the first evidence of new physics is expected to arise in the TeV energy range.  A high-energy, high-luminosity collider such as the muon collider will allow High Energy Physics to be explored at energy frontiers beyond the reach of existing and proposed colliders.
It is also important to mention that in this work we only focus on investigating the Higgs-strahlung production processes $\mu^{+}\mu^{-}\rightarrow (Z, Z') \to Zh_0, ZH_0$. However, producing Higgs bosons via vector boson fusion (VBF) is also possible. Given the high collision energy capabilities of the muon collider, the cross-sections for VBF processes could become significantly more important than Higgs-strahlung processes.  We are currently working on this complementary project, and it will be addressed in further work.

The outline of this paper is as follows:
% A  review of the BLHM is presented in Section~\ref{BLH}.
  Section~\ref{width} presents the decay widths of the $Z'$ boson in the BLHM. In Section~\ref{zh0H0}, we find the scattering amplitudes and cross-sections of the processes  $\mu^{+}\mu^{-}\rightarrow (Z, Z') \to Zh_0, Z H_0$.  Section~\ref{results} is devoted to our numerical results.  Finally, in Section~\ref{conclusions}, we present our conclusions. The Feynman rules involved in our calculations are provided in  Appendix~\ref{rulesF}.

\section{The total decay width of the $Z^{\prime}$ boson} \label{width}

In this section, we determine the total decay width of the $Z'$ boson, which we need to calculate the cross-section of the Higgs-strahlung processes. In the context of the BLHM, the main decay channels of the $Z'$ gauge boson are $Z' \to f \bar f$ ($f=t, b, T, T_5, T_6,  T^{2/3}, T^{5/3}, B$),  $Z' \to W^+W^-$, $Z' \to Zh_0$, and $Z' \to ZH_0$. It is worth mentioning that in the BLHM there is no $Z'H^{+}H^{-}$ coupling. Thus, the total decay width $\Gamma_{Z'}$ of the $Z'$  boson can be estimated as follows

\begin{eqnarray} \label{gammaTot}
    \Gamma_{Z'} &=&\sum_{f} \Gamma_{f\bar f}
    + \Gamma_{WW}+ \Gamma_{Zh_0}+ \Gamma_{ZH_0}.
\end{eqnarray}

\noindent We provide below the analytical expressions for the partial decay widths of the $Z'$ boson  involved in Eq.~(\ref{gammaTot}),

\begin{eqnarray} \label{gammaff}
\Gamma(Z' \to f\bar f) &=& \frac{N_c  m_{Z'}}{4 \pi}  \sqrt{1-\frac{4m^2_f}{m^2_{Z'}}} \bigg[ \left(g^{Z'ff}_V\right)^2
\left( 1+2 \frac{m^2_f}{m^2_{Z'}}  \right)  \nonumber \\
&+&  \left(g^{Z'ff}_A\right)^2 \left(1-4\frac{m^2_f}{m^2_{Z'}}\right) \bigg],
\end{eqnarray}

\begin{eqnarray}
    \Gamma(Z' \to W^+ W^-) = &-&\frac{m_{Z'}}{16\pi}\left(\frac{gc_W v^2 x_s}{f^2+F^2}\right)^2 \sqrt{1-\frac{4m_W^2}{m_{Z'}^2}} \left[ 1 + 12\left(\frac{m_W^2}{m_{Z'}^2}\right) \right. \nonumber \\
    &+& \left. 8\left(\frac{m_{Z'}^2}{m_W^2}\right) - \frac{9}{4}\left(\frac{m_{Z'}^4}{m_W^4}\right) \right],
\end{eqnarray}

\begin{eqnarray}
    \Gamma(Z' \to Zh_0) &=& \frac{g_{Z'Zh_0}^2}{16\pi m_{Z'}} \sqrt{\left[ 1- \left( \frac{m_{h_0} + m_Z}{m_{Z'}} \right)^2 \right] \left[ 1- \left( \frac{m_{h_0} - m_Z}{m_{Z'}} \right)^2 \right]} \left[ \frac{5}{2} + \frac{1}{4}\left( \frac{m_Z^2}{m_{Z'}^2} \right) \right. \nonumber \\
    &+& \left. \frac{1}{4}\left( \frac{m_{Z'}^2}{m_{Z}^2} \right) - \frac{1}{2}\left( \frac{m_{h_0}^2}{m_{Z'}^2} \right) - \frac{1}{2}\left( \frac{m_{h_0}^2}{m_{Z}^2} \right)  + \frac{1}{4}\left( \frac{m_{h_0}^2}{m_{Z'}^2} \right)\left( \frac{m_{h_0}^2}{m_{Z}^2} \right)  \right],
\end{eqnarray}

\begin{eqnarray}
\label{gammaZH}
    \Gamma(Z' \to ZH_0) &=& \frac{g_{Z'ZH_0}^2}{16\pi m_{Z'}} \sqrt{\left[ 1- \left( \frac{m_{H_0} + m_Z}{m_{Z'}} \right)^2 \right] \left[ 1- \left( \frac{m_{H_0} - m_Z}{m_{Z'}} \right)^2 \right]} \left[ \frac{5}{2} + \frac{1}{4}\left( \frac{m_Z^2}{m_{Z'}^2} \right) \right. \nonumber \\
    &+& \left. \frac{1}{4}\left( \frac{m_{Z'}^2}{m_{Z}^2} \right) - \frac{1}{2}\left( \frac{m_{H_0}^2}{m_{Z'}^2} \right) - \frac{1}{2}\left( \frac{m_{H_0}^2}{m_{Z}^2} \right)  + \frac{1}{4}\left( \frac{m_{H_0}^2}{m_{Z'}^2} \right)\left( \frac{m_{H_0}^2}{m_{Z}^2} \right)  \right],
\end{eqnarray}

\noindent where $N_c$ is the color factor ($N_c=1$ for leptons and $N_c=3$ for quarks), $g^{Z'ff}_V$ and $g^{Z'ff}_A$ are the vector and axial-vector coupling constants of the $Z'$ boson with the fermions (see Appendix~\ref{rulesF}~\cite{Gutierrez-Rodriguez:2023sxg}), and  $g_{Z'Zh_0}$ and $g_{Z'ZH_0}$ denote the effective couplings of the $Z^{\prime}$ and $Z$ bosons to the Higgs bosons $h_0$ and $H_0$ whose explicit expressions are given in Appendix~\ref{rulesF}. On the other hand,

\begin{equation}
    x_s= \frac{1}{2\, c_W} s_g c_g (s^2_{g} - c^2_{g}).
\end{equation}

\section{The Higgs-strahlung processes $\mu^+ \mu^- \to (Z,Z') \to Zh_0, Z H_0$ in the BLHM} \label{zh0H0}

\subsection{ Higgs-strahlung production $\mu^+\mu^{-} \to Zh_0$ }

\begin{figure}[H] 
\center
\subfloat[]{\includegraphics[width=6.5cm]{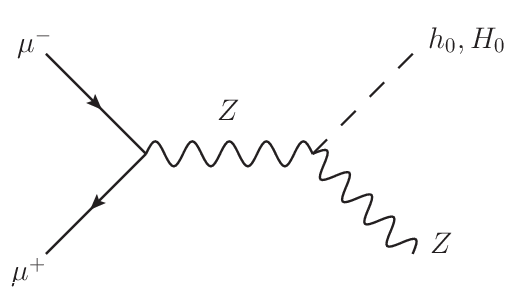}}
\subfloat[]{\includegraphics[width=6.5cm]{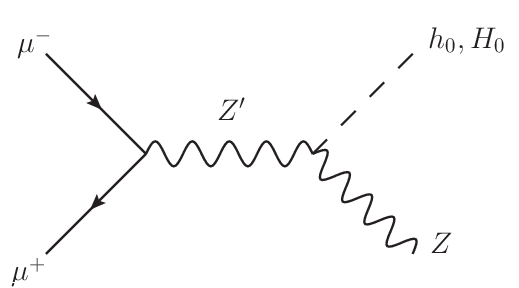}}
\caption{\label{strahlung}  Feynman diagrams  for the Higgs-strahlung production processes a) $\mu^{+}\mu^{-}\to Z \to Zh_0, ZH_0$ and b) $\mu^{+}\mu^{-}\to  Z' \to Zh_0, ZH_0$
in the BLHM.}
\end{figure}

The Feynman diagrams contributing to the Higgs-strahlung production processes $\mu^+\mu^{-} \to (Z, Z') \to Zh_0$ are shown in Fig.~\ref{strahlung}. The respective scattering amplitudes are represented by Eqs.~(\ref{MZ}) and~(\ref{MZp}),

\begin{eqnarray} 
{\cal M}_Z (\mu^+\mu^{-}  \to Zh_0) &=& g_{ZZh_0} \Bigl[ \bar v(p_1)\gamma^\mu (g^{Z\mu \mu}_V-g^{Z\mu \mu}_A\gamma_5)u(p_2) \Bigr ]
 \Bigl[ \frac{(-g_{\mu\nu} + p_{\mu}p_{\nu}/m^{2}_{Z})}{(p_{1}+p_{2})^{2}-m^{2}_{Z}+i m_{Z} \Gamma_{Z} }\Bigr] \nonumber \\
 &\times &  \epsilon^\nu_\lambda(Z),\label{MZ}\\
{\cal M}_{Z'} (\mu^+\mu^{-} \to Zh_0) &=&  g_{ZZ'h_0} \Bigl[ \bar v(p_1)\gamma^\mu (g^{Z\prime \mu \mu}_V-g^{Z\prime \mu \mu}_A\gamma_5)u(p_2) \Bigr ]
\Bigl[ \frac{(-g_{\mu\nu} + p_{\mu}p_{\nu}/m^{2}_{Z'})}{(p_{1}+p_{2})^{2}-m^{2}_{Z'}+i m_{Z^{\prime}} \Gamma_{Z'}}\Bigr]\nonumber \\
 &\times&  \epsilon^\nu_\lambda(Z), \label{MZp}
\end{eqnarray}

\noindent  where $\epsilon^\nu_\lambda(Z)$  represents the polarization vector of the $Z$ boson, while $p_{\mu}=(p_1+ p_2)_{\mu}$ is the four-momentum of the mediator particle. 
 The vector and vector-axial coupling constants of the $Z$ or $Z^{\prime}$ boson are given in Appendix~\ref{rulesF}.

We calculate from the transition amplitudes, Eqs.~(\ref{MZ}) and~(\ref{MZp}), the total cross-section $ \sigma^{Zh_0}_{T}$ for the processes $\mu^+ \mu^- \to (Z,Z') \to Zh_0$,

\begin{equation}
\label{XSTotEq}
    \sigma^{ Zh_0 }_{T}= \sigma^{ Zh_0 }_Z + \sigma^{ Zh_0 }_{Z'} + \sigma^{ Zh_0 }_{ZZ'},
\end{equation}

\noindent with

\begin{eqnarray}
 \sigma^{ Zh_0 }_{Z} &=& \frac{\sqrt{\lambda}}{192\pi} \left(\frac{(g^{Z\mu \mu}_V)^2 + (g^{Z\mu \mu}_A)^2}{m_Z^2 s^2}\right) \left(\frac{g_{ZZh_0}^2}{(s-m_{Z}^2)^2+(m_{Z}\Gamma_{Z})^2}\right) \left(12m_Z^2s + \lambda \right), \label{SZ} \\
 \sigma^{ Zh_0 }_{Z'} &=& \frac{\sqrt{\lambda}}{192\pi} \left(\frac{(g^{Z'\mu \mu}_V)^2 + (g^{Z'\mu \mu}_A)^2}{m_Z^2 s^2}\right) \left(\frac{g_{Z'Zh_0}^2}{(s-m_{Z'}^2)^2+(m_{Z'}\Gamma_{Z'})^2}\right) \left(12m_Z^2s + \lambda \right), \label{SZp} \\
   \sigma^{ Zh_0 }_{ZZ'} &=& \sqrt{\lambda} \left( \frac{g_{ZZh_0} g_{Z'Zh_0}}{96\pi} \right) \left(\frac{g^{Z\mu \mu}_V g^{Z'\mu \mu}_{V} + g^{Z\mu \mu}_A g^{Z'\mu \mu}_A}{m_Z^2 s^2} \right) \left(12m_Z^2s + \lambda\right)  \nonumber \\
     &\times& \frac{(s - m_Z^2)(s - m_{Z'}^2) + (m_Z \Gamma_Z)(\Gamma_{Z'}m_{Z'})}{((m_Z^2 - s)^2 + (m_Z\Gamma_Z)^2)((m_{Z'}^2 -
      s)^2 + (m_{Z'}\Gamma_{Z'})^2)}, \label{SZZp}
\end{eqnarray}

\noindent where $\sigma^{ Zh_0 }_Z$ and  $\sigma^{ Zh_0 }_{Z'}$  are the cross-sections of processes $\mu^+ \mu^- \to Z \to Zh_0$ and  $\mu^+ \mu^- \to Z' \to Zh_{0}$, respectively. The $\sigma^{ Zh_0 }_{ZZ'}$ cross-section represents the interference term between the $Z$ and $Z'$ bosons.
In Eqs. (\ref{SZ})-(\ref{SZZp}), $\sqrt{s}$ is the center-of-mass energy  and $\lambda$ is the usual two-particle phase space function,

\begin{equation}
    \lambda(s,m_{Z},m_{h_0}) = (s-m_Z^2-m_{h_0}^2)^2 - 4m_Z^2 m_{h_0}^2.
\end{equation}

It is appropriate to mention that Eq.~(\ref{XSTotEq}), which corresponds to the total cross-section in the context of the BLHM, reproduces the cross-section for the process $\mu^+ \mu^- \to Z \to Zh_0$ obtained in the SM scenario~\cite{ParticleDataGroup:2024cfk,Ellis,Ioffe,Lee,Bjorken,Barger}. This is reached in the decoupling limit of the new physics scales, i.e., $(f, F) \to \infty$.

To quantify the combined effects of the free parameters  $\sqrt{s}$, $f$ and $F$ of the BLHM, we define the relative correction
for the total  cross-section as follows,

\begin{equation} \label{relative}
    \frac{\delta\sigma^{BLHM}}{\sigma^{SM}} = \frac{\sigma^{Z h_0}_{T} (\sqrt{s},  f, F) -\sigma^{SM}(\sqrt{s})}{\sigma^{SM}(\sqrt{s})},
\end{equation}

\noindent where $\sigma^{SM}(\sqrt{s})$ represent the cross-section of the SM and $\sigma^{Z h_0}_{T} (\sqrt{s},  f, F)$ represent
the cross-section in the presence of interactions of the BLHM.

\subsection{ Higgs-strahlung production $\mu^+\mu^{-} \to Z H_0$ }

We determine the scattering amplitudes and cross-sections of Higgs-strahlung production $\mu^+\mu^- \to (Z, Z') \to ZH_0$. The transition amplitudes are obtained from the Feynman diagrams contributing to the process $\mu^+\mu^- \to ZH_0$ (see Fig.~\ref{strahlung}), which are given by

\begin{eqnarray}
{\cal M}_Z (\mu^+\mu^{-}  \to ZH_0) &=&  g_{ZZH_0} \Bigl[ \bar v(p_1)\gamma^\mu (g^{Z \mu \mu}_V-g^{Z \mu \mu}_A\gamma_5)u(p_2) \Bigr ]
\Bigl[ \frac{(-g_{\mu\nu} + p_{\mu}p_{\nu}/m^{2}_{Z})}{(p_{1}+p_{2})^{2}-m^{2}_{Z}+ i m_{Z} \Gamma_{Z}} \Bigr] \nonumber \\
&\times &  \epsilon^\nu_\lambda(Z), \label{SZH0} \\
{\cal M}_{Z'} (\mu^+\mu^{-}  \to Z H_0) &=&   g_{ZZ'H_0} \Bigl[ \bar v(p_1)\gamma^\mu (g^{Z' \mu \mu}_V-g^{Z' \mu \mu}_A\gamma_5)u(p_2) \Bigr ]
\Bigl[\frac{(-g_{\mu\nu} + p_{\mu}p_{\nu}/m^{2}_{Z'})}{(p_{1}+p_{2})^{2}-m^{2}_{Z'}+ i m_{Z'}\Gamma_{Z'}} \Bigr]  \nonumber \\
&\times & \epsilon^\nu_\lambda(Z). \label{SZPH0}
\end{eqnarray}

\noindent Using Eqs.~(\ref{SZH0}) and~(\ref{SZPH0}), we calculate the total cross-section for the process $\mu^+ \mu^- \to (Z,Z') \to ZH_0$, which can be written in the following compact form,

\begin{equation} \label{ZH0tot}
    \sigma^{ZH_0}_{T}= \sigma^{ZH_0}_Z + \sigma^{ZH_0}_{Z'} + \sigma^{ZH_0}_{ZZ'},
\end{equation}

\noindent where

\begin{eqnarray}
   \sigma^{ZH_0}_{Z} &=& \frac{\sqrt{\lambda}}{192\pi} \left(\frac{(g^{Z\mu \mu}_V)^2 + (g^{Z\mu \mu}_A)^2 }{m_Z^2 s^2}\right) \left(\frac{g_{ZZH_0}^2}{(s-m_{Z}^2)^2+(m_{Z}\Gamma_{Z})^2}\right) \left(12m_Z^2s + \lambda \right), \label{ZH0-Z}\\
\sigma^{ZH_0}_{Z'} &=& \frac{\sqrt{\lambda}}{192\pi} \left(\frac{(g^{Z' \mu \mu}_V)^2 + (g^{Z' \mu \mu}_A)^2}{m_Z^2 s^2}\right) \left(\frac{g_{Z'ZH_0}^2}{(s-m_{Z'}^2)^2+(m_{Z'}\Gamma_{Z'})^2}\right) \left(12m_Z^2s + \lambda \right), \label{ZH0-Zp} \\
 \sigma^{ZH_0}_{ZZ'} &=& \sqrt{\lambda} \left( \frac{g_{ZZH_0} g_{Z'ZH_0}}{96\pi} \right) \left(\frac{ g^{Z\mu \mu}_V g^{Z'\mu \mu}_{V} + g^{Z\mu \mu}_A g^{Z'\mu \mu}_A }{m_Z^2 s^2} \right) \left(12m_Z^2s + \lambda\right)  \nonumber \\
     &\times& \frac{(s - m_Z^2)(s - m_{Z'}^2) + (m_Z \Gamma_Z)(\Gamma_{Z'}m_{Z'})}{((m_Z^2 - s)^2 + (m_Z\Gamma_Z)^2)((m_{Z'}^2 - s)^2 + (m_{Z'}\Gamma_{Z'})^2)}. \label{ZH0-mezcla}
\end{eqnarray}

\noindent In these expressions, the two-particle phase space function is given by

\begin{equation}
    \lambda(s,m_{Z},m_{H_0}) = (s-m_Z^2-m_{H_0}^2)^2 - 4m_Z^2 m_{H_0}^2.
\end{equation}

\section{Numerical results} \label{results}

In our numerical analysis of the Higgs-strahlung production processes $\mu^+\mu^-   \to Zh_0$ and $\mu^+\mu^-   \to  ZH_0$, various LHC measurements are used to constrain specific relevant parameters of the BLHM. Below, we summarise the different input parameters, searches, and measurements used for our analysis.

 The Yukawa couplings $y_i$ ($i=1,2,3$)~\cite{Godfrey:2012tf,Cruz-Albaro:2023pah,Cruz-Albaro:2022kty} are related to the generation of the new quark masses: $ T $, $ B $, $ T_5 $, $ T_6 $, $ T^{2/3} $, $ T^{5/3} $.
 These Yukawa couplings generate two study scenarios, which arise because in the region where $y_2\approx y_3$, the masses of the $T$ and $T_5$ states are degenerate~\cite{Godfrey:2012tf}. The two scenarios to which we refer are

\begin{itemize}
\item Scenario  {\bf a} ($y_2 > y_3$): $y_1=0.61$, $y_2=0.84$, and $y_3=0.35$~\cite{Cruz-Albaro:2023pah,Cruz-Albaro:2022kty,Cruz-Albaro:2022lks},
\item Scenario {\bf b} ($y_2 < y_3$): $y_1=0.61$, $y_2=0.35$, and $y_3=0.84$~\cite{Cruz-Albaro:2023pah,Cruz-Albaro:2022kty,Cruz-Albaro:2022lks}.
\end{itemize}

\noindent In the first scenario ($y_2 > y_3$), the mass separation between the new quarks $T_5$ and $T_6$ is relatively tiny and leads to the decays of $T_5$ being predominantly to SM particles. For the second scenario ($y_2 < y_3$), the mass separation between the $T_5$ and $T_6$ states is large, which increases the decay modes available for the $T_5$ quark through decay cascades to non-SM particles~\cite{Godfrey:2012tf,PhenomenologyBLH}. 
Because of the phenomenological implications of the first scenario, in this paper we explore the Higgs-strahlung productions $\mu^+\mu^-   \to Zh_0$ and $\mu^+\mu^-   \to  ZH_0$ in the $y_2 > y_3$ scenario. Other parameters involved in our calculations are discussed briefly below. 
%On the other hand, our choice in the values of $y_i$ is motivated by the perturbativity requirements~\cite{Altmannshofer:2010zt} and the fine-tuning measure~\cite{JHEP09-2010,Godfrey:2012tf}. Using numerical methods, from Eq.~(\ref{yt}), we randomize perturbative values of the Yukawa couplings $y_i$  by fixing $y_t$ through experimental measurement of the top quark mass ($m_t=172.69$ GeV~\cite{Workman:2022ynf})  and the Higgs boson mass ($ m_{h_0}=125.25 $ GeV~\cite{Workman:2022ynf}), and an appropriate choice of the other parameters involved ($m_{A_0}$,  $\tan \beta$, and others), as discussed below.  

\vspace{0.1cm}

% \noindent \textbf{ The pseudoscalar mass $A_{0}$}:
  \noindent  $\bf{m}_{A_{0}}$: 
  The mass of the  $A_0$ pseudoscalar is an input parameter of the BLHM. This parameter is set to 1000 GeV, which is consistent with current searches for new scalar bosons~\cite{ATLAS:2020gxx,CMS:2019ogx}. 
% According to data recorded by the ATLAS experiment at the LHC, it corresponds to an integrated luminosity of 139 $\text{fb}^{-1}$ from $pp$ collisions at center-of-mass energy 13 TeV were used to search for a heavy Higgs boson, $A_{0}$, decaying into $ZH_0$, where $H_0$ denotes another  Higgs boson with mass $m_{H_0}>125$ GeV.
\vspace{0.1cm}

 %\noindent \textbf{The mass of the Higgs boson  $H_{0}$}: 
 \noindent  $\bf{m_{H_{0}}}$:
 The mass of the Higgs boson $H_0$ is calculated from the mass of the $A_0$ pseudoscalar  and the values of $\tan\, \beta$ (the ratio of the vacuum expectation values (VEVs) of the two Higgs doublets)~\cite{Kalyniak:2013eva}:
 
 \begin{eqnarray}\label{massH0}
m^{2}_{H_{0}} &=& \frac{B_\mu}{\text{sin}\, 2\beta}+ \sqrt{\frac{B^{2}_{\mu}}{\text{sin}^{2}\, 2\beta} -2\lambda_0 B_\mu v^{2} \text{sin}\, 2\beta +\lambda^{2}_{0} v^{4} \text{sin}^{2}\, 2\beta  }
\end{eqnarray}
\noindent where
 \begin{eqnarray}
B_\mu &=&\frac{1}{2}(\lambda_0  v^{2} + m^{2}_{A_{0}}  )\, \text{sin}\, 2\beta,\\
\lambda_0 &=& \frac{m^{2}_{h_{0}}}{v^{2}}\Big(\frac{  m^{2}_{h_{0}}- m^{2}_{A_{0}} }{m^{2}_{h_{0}}-m^{2}_{A_{0}} \text{sin}^{2}\, 2\beta }\Big).
\end{eqnarray}
\vspace{0.1cm}

\noindent  \textbf{The ratio of the VEVs $v_1$ and $v_2$ ($\tan \beta$)}:
The authors of Refs.~\cite{JHEP09-2010,Kalyniak:2013eva,Altmannshofer:2010zt} set lower and upper bounds on the parameter $\tan \beta$ which arises due to perturbativity requirements on the parameter $\lambda_0$.
 Thus, the range of values that  $\tan \beta$  could acquire is set according to the following equation
 
 \begin{eqnarray}\label{parametros}
1 < &\text{tan} & \beta  <  \sqrt{ \frac{2+2 \sqrt{\big(1-\frac{m^{2}_{h_0} }{m^{2}_{A_0}} \big) \big(1-\frac{m^{2}_{h_0} }{4 \pi v^{2}}\big) } }{ \frac{m^{2}_{h_0}}{m^{2}_{A_0}} \big(1+ \frac{m^{2}_{A_0}- m^{2}_{h_0}}{4 \pi v^{2}}  \big) } -1 }.
\end{eqnarray}
 
\noindent For $m_{A_{0}}=1000$ GeV, it is obtained that $1 < \tan \beta < 10.45$.  Consistently, in this work, we have chosen $\tan \beta=3$ and $\tan \beta=6$~\cite{Cruz-Albaro:2023pah,Cruz-Albaro:2022kty,Cruz-Albaro:2022lks} to carry out our numerical analysis of the production of $Z h_0$ and $Z H_0$ at a future muon collider. 
\vspace{0.1cm}

\noindent \textbf{Gauge couplings}:
The  gauge couplings $g_{A}$ and  $g_{B}$ can be parametrized in terms of the mixing angle $\theta_{g}$ and the electroweak gauge coupling: 
$\tan \theta_{g}=g_{A}/g_{B}$  and $g=g_{A} g_{B}/\sqrt{g^{2}_{A}+ g^{2}_{B}} $. 
For our study, it is assumed that the gauge coupling $g_B= \frac{1}{2} g_A$, implies that $g_A= \sqrt{5}g$. 
Another possible study scenario arises when $g_{A} =g_{B}$; however, this project could be left for  later work.

\vspace{0.1cm}
\noindent \textbf{ Symmetry breaking scales ($f$, $F$)}:
The BLHM is characterized because it incorporates two different global symmetries that are broken into diagonal subgroups at different scales, $f$ and $F$.
With respect to the $f$ scale, certain limits on this parameter arise when considering fine-tuning constraints on the masses of heavy quarks, as well as experimental constraints on the production of quarks: $f\in [700,3000]$ GeV~\cite{Kalyniak:2013eva,Godfrey:2012tf}.
On the  other hand, the energy scale $F$ acquires sufficiently large values compared to the $f$ scale. The purpose is to ensure that the new gauge bosons are much heavier than the new quarks, thus, $F > 3000$ GeV~\cite{JHEP09-2010,Kalyniak:2013eva}.

  One of the motivations for building the BLHM, is to avoid fine-tuning the Higgs potential. In this way, scenarios {\bf a} and {\bf b} mentioned above provide realistic values of the Yukawa couplings as they  minimize the fine-tuning constraints. 
  From  Eq.~(\ref{fine-tuning}), 
  
\begin{eqnarray} \label{fine-tuning}
\Psi= \frac{27 f^2}{8 \pi^{2} \lambda_0 v^{2} \cos^{2} \beta } \frac{ | y_{1}|^{2} | y_{2}|^{2} | y_{3}|^{2}  }{| y_{2}|^{2}- | y_{3}|^{2} }\, \text{log} \left( \frac{| y_{1}|^{2} + | y_{2}|^{2}}{| y_{1}|^{2} + | y_{3}|^{2}} \right),
\end{eqnarray}
   
\noindent we determine the measure of the fine-tuning for certain values of the scale $f$.  In Tables~\ref{ajuste1} and~\ref{ajuste1-tan6}, we show a measure of the fine-tuning when the energy scale $f$ takes on values such as 1.0, 1.5, 2.0, 2.5, and 3.0 TeV.
When $\tan\, \beta=3$ (see Table~\ref{ajuste1}), the size of the fine-tuning for $f=1.0$ TeV is $\Psi=0.54$, which indicates that there is no fine-tuning in the BLHM~\cite{JHEP09-2010,PhenomenologyBLH}. The absence of fine-tuning prevails up to $\Psi=2.2$, that is, for values of the $f$ scale close to 2 TeV.  The fine-tuning starts to become relevant for $f > 2.1$ TeV.
With respect to Table~\ref{ajuste1-tan6} generated for $\tan\, \beta=6$, the absence of fine-tuning only prevails for points close to $f=1$ TeV, for $f>1.2$ TeV the model needs to be fine-tuned.

\begin{table}[H]
\begin{center}
\caption{A measure of the fine-tuning in the BLHM for some values of the $f$ scale. The values generated for $\Psi$ are obtained by setting $\tan\, \beta=3$.}
\begin{tabular}{|c|c|}
\hline
\multicolumn{2}{|c|}{ $\tan\, \beta=3$} \\
\hline
\hspace{0.6cm} $f$ [TeV] \hspace{0.6cm} & \hspace{1cm} $\Psi$ \hspace{1cm} \\\hline
$1.0$ & $0.54$\\\hline
$1.5$ & $1.21$\\\hline
$2.0$ & $2.16$\\\hline
$2.5$ & $3.37$\\\hline
$3.0$ & $4.85$\\\hline
 \end{tabular}
 \label{ajuste1}
 \end{center}
 \end{table}

 \begin{table}[H]
\begin{center}
\caption{A measure of the fine-tuning in the BLHM for some values of the $f$ scale. The values generated for $\Psi$ are obtained by setting $\tan\, \beta=6$.}
\begin{tabular}{|c|c|}
\hline
\multicolumn{2}{|c|}{ $\tan\, \beta=6$} \\
\hline
\hspace{0.6cm} $f$ [TeV] \hspace{0.6cm} & \hspace{1cm} $\Psi$ \hspace{1cm} \\\hline
$1.0$ & $1.99$\\\hline
$1.5$ & $4.49$\\\hline
$2.0$ & $7.98$\\\hline
$2.5$ & $12.47$\\\hline
$3.0$ & $17.96$\\\hline
 \end{tabular}
 \label{ajuste1-tan6}
 \end{center}
 \end{table}
  
As a summary, we provide in Table~\ref{parametervalues} the values assigned to the parameters involved in our calculation.

\begin{table}[H]
\caption{Values assigned to the parameters involved in our numerical analysis at the BLHM.
\label{parametervalues}}
\centering
\begin{tabular}{|c | c | c |}
\hline
\hspace{0.3cm} $ \textbf{Parameter} $ \hspace{0.5cm}  &  \hspace{1.2cm}  $\textbf{Value} $ \hspace{1.2cm}  &  \hspace{0.5cm}   $ \textbf{Reference} $ \hspace{0.3cm} \\
\hline
\hline
$ m_{h_{0}}  $  &   $ 125.20\  \text{GeV} $ &  \cite{ParticleDataGroup:2024cfk}  \\
\hline
$ m_{A_{0}}  $  &   $ 1000\  \text{GeV} $ &   \cite{ATLAS:2020gxx,CMS:2019ogx}  \\
\hline
$ \tan \beta $  &    $ 3, 6 $  &  \cite{Cruz-Albaro:2023pah,Cruz-Albaro:2022kty,Cruz-Albaro:2022lks} \\
\hline
$ m_{H_{0}}  $  &   $ 1015\  \text{GeV} $ &    \\
\hline
$ g_{A} $  &   $ \sqrt{5}\, g $ &  \\
\hline
$\Gamma_Z$  &  $2.4955\pm 0.0023$ GeV  & \cite{ParticleDataGroup:2024cfk} \\
\hline
$ f $  &  $ [1000, 3000]\   \text{GeV} $ &  \cite{JHEP09-2010,Godfrey:2012tf,Cruz-Albaro:2023pah, Cruz-Albaro:2022kty,Cruz-Albaro:2022lks}  \\
\hline
$ F $  &   $ > 3000  \ \text{GeV} $ &  \cite{JHEP09-2010,Kalyniak:2013eva,Godfrey:2012tf,Cruz-Albaro:2023pah,Cruz-Albaro:2022kty,Cruz-Albaro:2022lks} \\
\hline
\end{tabular}
%\caption{BLHM contributions to $a^{W}_{t}$.
%\label{CN2}}
\end{table}

\subsection{$\Gamma_{Z^{\prime}} $}

\begin{figure}[H]
\center
\subfloat[]{\includegraphics[width=10.80cm]{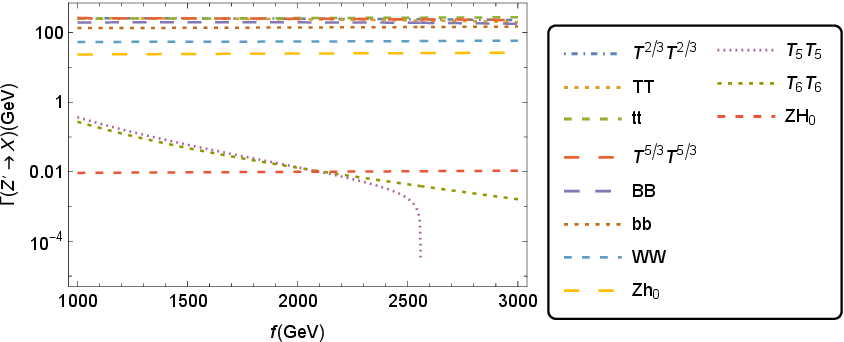}}\\
\subfloat[]{\includegraphics[width=10.80cm]{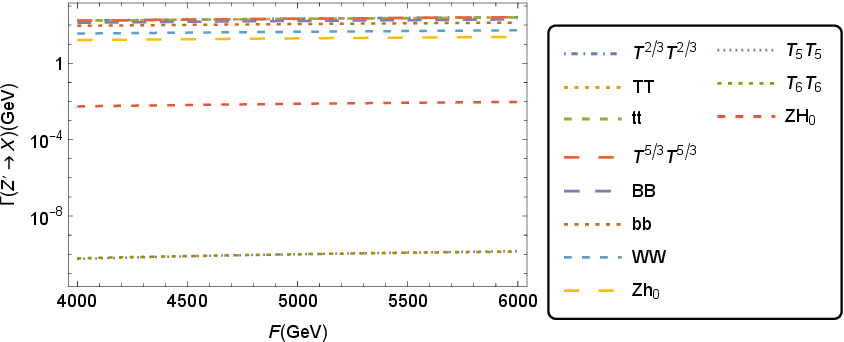}}
\caption{ \label{widhts} Decay widths for the processes  $Z' \to X$  where $X= t\bar t,  T\bar T, T_5\bar T_5,$ $ T_6\bar T_6, T^{2/3}\bar T^{2/3},
T^{5/3}\bar T^{5/3}, b\bar b, B\bar B, Z h_{0}, Z H_{0}, W W$. a) $\Gamma(Z^{\prime} \to X)$ as a function of the $f$ energy scale (with the fixed value of $F=6\, 000$ GeV). b) $\Gamma(Z^{\prime}  \to X)$ as a function of the $F$ energy scale (with the fixed value of $f=1\, 000$ GeV).
}
\end{figure}

Another of the essential input parameters involved in our study of the Higgs-strahlung production $\mu^+\mu^- \to (Z, Z')  \to Zh_0, ZH_0$ is the total decay width of the $Z'$ boson ($\Gamma_{Z^{\prime}}$), which has a dependence on the two energy scales, $f$ and $F$,  these represent the scales of the new physics in the BLHM. 
In this subsection, we analyze the different contributions that receive $\Gamma_{Z^{\prime}}$, and we also discuss the behavior of the partial widths $\Gamma(Z' \to X)$ when the  $f$ scale takes values from $1 000$ to $3 000$ GeV while keeping the  $F$ scale fixed, and when $F$ varies from $4 000$ to $6 000$ GeV while fixing $f$ (see Fig.~\ref{widhts}). In the left plot of Fig.~\ref{widhts}, we show the evolution of $\Gamma(Z' \to X)$ vs. $f$; these curves are generated by setting $F=6 000$ GeV. In this scenario, the main partial contributions are generated by the decays $Z^{\prime} \to T^{2/3}\bar T^{2/3} $ and $Z^{\prime} \to  T\bar T $. These provide the dominant and subdominant numerical contributions:  $\Gamma(Z' \to  T^{2/3}\bar T^{2/3})=[262.43, 256.61]$ GeV and  $\Gamma(Z' \to T\bar T)=[259.70, 249.83]$ GeV, respectively. This occurs while $f\in [1000, 1650]$ GeV, outside this interval, $ \Gamma(Z' \to  t\bar t)$ becomes dominant.
On the opposite side, the most suppressed contribution is given by the  $Z' \to  Z H_0$ decay: $\Gamma(Z' \to  Z H_0)=[9.27 \times 10^{-3}, 1.07\times 10^{-2}]$ GeV over the whole analysis interval of the $f$ scale. With respect to the remaining curves, $\Gamma(Z' \to  T^{5/3}\bar T^{5/3}) \approx \Gamma(Z' \to   B\bar B) \approx \Gamma(Z' \to   b\bar b) \in [2.60, 1.35]\times 10^{2}$ GeV, $\Gamma(Z' \to WW) \approx \Gamma(Z' \to   Z h_0)  \in [6.0, 2.3] \times 10^{1}$ GeV, and 
$\Gamma(Z' \to  T_5\bar T_5) \approx \Gamma(Z' \to   T_6\bar T_6) \in [10^{-1}, 10^{-3}]$ GeV. 
On the other hand, in the right plot of Fig.~\ref{widhts}, we can appreciate the behavior of $\Gamma(Z' \to X)$ vs. $F$, these curves have been generated for $f=1000$ GeV. In this case, the most significant contributions are given by the decays  $Z' \to T^{2/3}\bar T^{2/3}$  and  $Z' \to  T\bar T$: 
 $\Gamma(Z' \to T^{2/3}\bar T^{2/3})=[172.99, 262.43]$ GeV and  $\Gamma(Z' \to  T\bar T)=[169.08, 259.70]$ GeV when the $F$ scale obtains values in the interval from 4000 to 6000 GeV. On the contrary, the minor contribution is led by the process $Z' \to Z H_0$,  $\Gamma(Z' \to  Z H_0)=[5.39,9.27]\times 10^{-3}$ GeV. The other curves approximately take on values in the following ranges:  $\Gamma(Z' \to  t\bar{t})\approx \Gamma(Z' \to  T^{5/3}\bar T^{5/3})  \approx \Gamma(Z' \to  B\bar{B}) \in [1.0, 2.5] \times 10^{2} $ GeV,  $\Gamma(Z' \to  b\bar{b})\approx \Gamma(Z' \to  W W)  \approx \Gamma(Z' \to  Z h_0) \in [0.9\times 10^{2}, 1.0 \times 10^{1}] $ GeV, and $\Gamma(Z' \to  T_5\bar T_5)\approx \Gamma(Z' \to  T_6 \bar T_6) \sim 10^{-1}$ GeV.
 From the above,  $\Gamma(Z' \to  X)$ shows a strong sensitivity to variations in the scales of the new physics ($f$ and $F$). 
  The dependence of  $\Gamma(Z' \to  X)$ on $F$ is more pronounced with respect to $f$, and all curves show increasing behavior (see Fig.~\ref{widhts}(b)). 
In addition, the dominance of the fermionic decay modes of $Z'$ ($Z^{\prime} \to T^{2/3}\bar T^{2/3}, T\bar{T}, t \bar{t}$) leads to the assumption that the discovery channels should be related to the new QCD fermions.

\subsection{$\text{Br}\left(Z^{\prime} \to X \right) $}

\begin{figure}[H]
\center
\subfloat[]{\includegraphics[width=10.80cm]{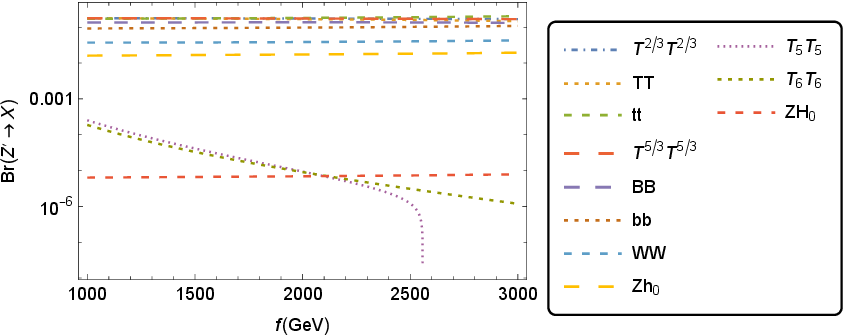}}\\
\subfloat[]{\includegraphics[width=10.80cm]{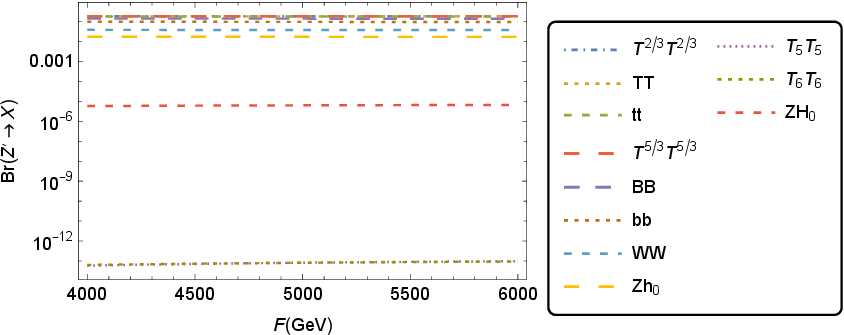}}
\caption{ \label{branching} The branching ratios for the processes $Z' \to X$  where $X= t\bar t,  T\bar T, T_5\bar T_5,$ $ T_6\bar T_6, T^{2/3}\bar T^{2/3},
T^{5/3}\bar T^{5/3}, b\bar b, B\bar B, Z h_{0}, Z H_{0}, W W$. a) $\text{Br}(Z^{\prime} \to X)$ as a function of the $f$ energy scale (with the fixed value of $F=6\, 000$ GeV). b) $\text{Br}(Z^{\prime}  \to X)$ as a function of the $F$ energy scale (with the fixed value of $f=1\, 000$ GeV).
}
\end{figure}

We now present our results on the branching ratios of the  $Z'$  gauge boson as a function of the scales of new physics ($f$ or $F$). 
We first discuss the behavior observed in Fig.~\ref{branching}(a).   In this figure, we plot Br$(Z' \to X)$ vs. the $f$ scale while fixing the second $F$ scale at 6000 GeV. The curve that provides the dominant contribution is given by the $Z'\to T^{2/3}\bar{T}^{2/3}$ decay, its associated branching ratio is Br$(Z' \to T^{2/3} \bar{T}^{2/3})=[1.80,1.68]\times 10^{-1}$ when $f\in [1\, 000, 3\, 000]$ GeV.
On the opposite side, we find that the $Z' \to ZH_0$ decay provides the most suppressed contribution,  Br$(Z' \to Z H_0)=[6.35, 7.85]\times 10^{-6}$.
As far as the remaining branching ratios are concerned, these acquire values of Br$(Z' \to T\bar{T})\sim $ Br$(Z' \to t \bar{t})\sim $ Br$(Z' \to T^{5/3}\bar{T}^{5/3})\sim $ Br$(Z' \to B\bar{B})\sim 10^{-1} $, Br$(Z' \to b\bar{b})\sim $ Br$(Z' \to Z h_0)\sim $ Br$(Z' \to W W)\sim 10^{-1}-10^{-2}$, and Br$(Z' \to T_5\bar{T}_5)\sim \text{Br}(Z' \to T_6\bar{T}_6)\sim 10^{-4}-10^{-6}$.
Concerning Fig.~\ref{branching}(b), here we explore the behavior of Br$(Z' \to X)$ vs. the $F$ scale in the interval from 4000 to 6000 GeV, and all the curves shown in the figure have been generated with the fixed value of $f=1\, 000$ GeV. 
In this scenario, we can appreciate that the curve that provides the slightly more significant contribution is derived from the $Z' \to T^{2/3} \bar{T}^{2/3}$ decay, 
Br$(Z' \to T^{2/3}\bar{T}^{2/3})=[1.79,1.80]\times 10^{-1}$. On the other hand, the smallest contribution is given by Br$(Z' \to Z H_0)=[5.58, 6.35]\times 10^{-6}$. 
The decays  $Z' \to    T\bar T$,  $Z' \to  t\bar t$,  $Z' \to T^{5/3}\bar T^{5/3}$ and  $Z' \to  B\bar B$ also generate branching ratios with values of the same order of magnitude than the main contribution although slightly smaller.
Complementarily, the other branching ratios acquire values of Br$(Z' \to  b\bar b)\sim $  Br$(Z' \to   W W)\sim $ Br$(Z' \to  Z h_0)\sim 10^{-2}$,  Br$(Z' \to  T_5\bar T_5)\sim 10^{-4}$ and Br$(Z' \to  T_6\bar T_6)\sim 10^{-5}$.
In conclusion, the values obtained by the branching ratios Br$(Z' \to X)$ do not show appreciable changes as the $F$ scale increases to 6000 GeV, as shown in the corresponding figure.  Br$(Z' \to X)$ slightly depends on the $F$ scale compared to the $f$ scale.
The new heavy quarks and the top quark of the SM are the most likely decays of the new gauge boson $Z'$.

\subsection{ Higgs boson production $h_0$ in the BLHM} \label{ph0}

\begin{figure}[H]
\center
\includegraphics[width=0.7\textwidth]{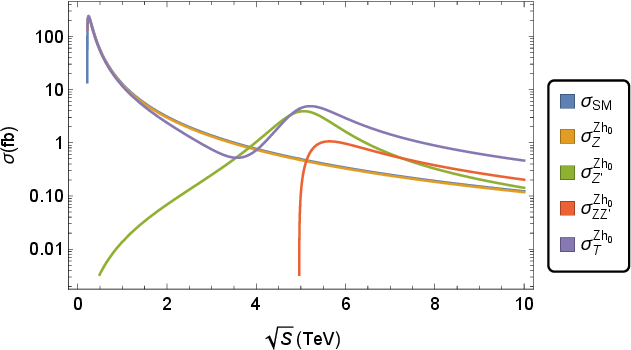}
\caption{The cross-section of the process  $\mu^{+}\mu^{-}\to (Z, Z') \to Zh_0$   as a function of $\sqrt{s}$.  The curves are generated for   $f = 1000$ GeV and $F=6000$ GeV (for $m_{Z'} = 5\, 200$ GeV), and correspond to  $\sigma_{SM}$, $\sigma^{Z h_0}_Z$ (Eq.~(\ref{SZ})), $\sigma^{Z h_0}_{Z'}$ (Eq. (\ref{SZp})), $\sigma^{Z h_0}_{ZZ'}$ (Eq. (\ref{SZZp})), and  $\sigma^{Z h_0}_{T} $ (Eq. (\ref{XSTotEq})).} \label{sigma-zh0}
\end{figure}

We investigate the Higgs-strahlung production process $\mu^{+} \mu^{-} \to Zh_0$ at the future muon collider and calculate the BLHM predictions on the $\sigma^{Z h_0}_{T}\left(\sqrt{s}, f, F \right)$ cross-section. We scan the BLHM parameters by considering various experimental and theoretical constraints.
In Fig.~\ref{sigma-zh0}, we present our results for the total cross-section $\sigma^{Z h_0}_{T}\left(\mu^{+} \mu^{-} \to Zh_0 \right)$  where the resonant and non-resonant effects of the processes $\mu^{+} \mu^{-}\rightarrow (Z, Z') \to Zh_0$  are taken into account. We also show the different contributions received by $\sigma^{Z h_0}_{T}\left(\mu^{+} \mu^{-} \to Zh_0 \right)$ according to Eq.~(\ref{XSTotEq}): $\sigma^{Z h_0}_{Z}\left(\mu^{+} \mu^{-} \to Zh_0 \right)$, $\sigma^{Z h_0}_{Z'}\left(\mu^{+} \mu^{-} \to Zh_0 \right)$, and $\sigma^{Z h_0}_{Z Z'}\left(\mu^{+} \mu^{-} \to Zh_0 \right)$. Complementarily, we plot in this same scenario the contribution of the SM represented by $\sigma_{SM}\left(\mu^{+} \mu^{-} \to Zh_0 \right)$.
From Fig.~\ref{sigma-zh0}, we can appreciate that the behavior of the cross-section $\sigma_{SM}\left(\mu^{+} \mu^{-} \to Zh_0 \right)$ very closely resembles the behavior of the curve represented by $\sigma^{Z h_0}_{Z}\left(\mu^{+} \mu^{-} \to Zh_0 \right)$. The latter corresponds to the cross-section with the $Z$ boson exchange in the context of the BLHM. In this particular case, the contribution of new physics is almost negligible.
 Concerning the other curves, $\sigma^{Z h_0}_{Z'}$ and $ \sigma^{Z h_0}_{T}$, these obtain an increase in the cross-section for large values of the center-of-mass energy, reaching their maximum value at the resonance of the  $Z'$ gauge boson, i.e., when $\sqrt{s}\approx 5\, 200 $ GeV:
 $\sigma^{Z h_0}_{Z'}\left(\mu^{+} \mu^{-} \to Zh_0 \right)=3.72$ fb, and $\sigma^{Z h_0}_{T}\left(\mu^{+} \mu^{-} \to Zh_0 \right)=4.04$ fb.  For this benchmark, $\sigma^{Z h_0}_{Z Z'}\left(\mu^{+} \mu^{-} \to Zh_0 \right)$  its contribution is $0.64$ fb.
It is also important to mention that in certain regions the total cross-section of the process $\mu^{+}\mu^{-} \to (Z, Z') \to Zh_0$ takes values smaller than the sum of the individually contributing processes, i.e. $\mu^{+}\mu^{-} \to Z \to Zh_0$ and $\mu^{+}\mu^{-} \to Z' \to Zh_0$. This effect is basically due to the negative interference between the channels  $\mu^{+}\mu^{-} \to Z \to Zh_0$ and $\mu^{+}\mu^{-} \to Z' \to Zh_0$. As well as the effect of the effective couplings $g_{ZZ'h_0}$ and $g_{ZZh_0}$, which contain positive and negative terms, as can be seen in Appendix~\ref{rulesF}.

\begin{figure}[H]
        \centering
        \includegraphics[width=0.8\textwidth]{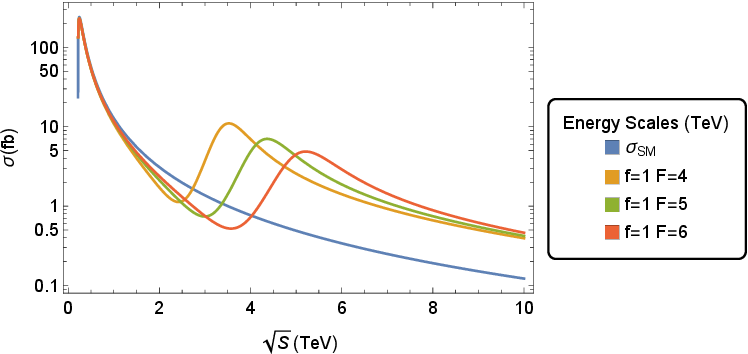}
        \caption{The total cross-section of the process $\mu^{+}\mu^{-}\to (Z, Z') \to Zh_0$   as a function of $\sqrt{s}$.
        The curves are generated for $f=1000$ GeV and $F=4 000$ GeV, $f=1 000$ GeV and $F=5000$ GeV, and $f=1 000$ GeV and $F=6 000$ GeV.}
        \label{sT-fF}
\end{figure}

We test through the total cross-section the effects that could be provided by the new physics scales, $f$ and $F$, on $\sigma^{Z h_0}_{T}\left(\mu^{+} \mu^{-} \to Zh_0 \right)$.  In this way, in Fig.~\ref{sT-fF} we show the different curves generated for $\sigma^{Z h_0}_{T}\left(\mu^{+} \mu^{-} \to Zh_0 \right)$  when the  $f$ and $F$ scales take specific fixed values while the center-of-mass energy $\sqrt{s}$ varies in the interval from 0 to 10 000 GeV.  In this figure, we can see that the curves corresponding to $\sigma^{Z h_0}_{T}\left(\mu^{+} \mu^{-} \to Zh_0 \right)$ decrease for large values of $\sqrt{s}$. 
We also note that the height of the resonance peaks for the $Z'$ boson changes depending on the value of the $F$ scale. For the plotted curves, $\sigma^{Z h_0}_{T}\left(\mu^{+} \mu^{-} \to Zh_0 \right)$ reaches its local maxima just at the resonance of the  $Z'$ gauge boson:
$\sigma^{Z h_0}_{T}\left(\sqrt{s}, 1\, 000 \ \text{GeV},  4\, 000\ \text{GeV} \right)=10.99$ fb, $\sigma^{Z h_0}_{T}\left(\sqrt{s}, 1\, 000 \ \text{GeV},  5\, 000\ \text{GeV} \right)=7.02$ fb  and $\sigma^{Z h_0}_{T}\left(\sqrt{s}, 1\, 000 \ \text{GeV},  6\, 000\ \text{GeV} \right)=4.87$ fb for $m_{Z'} \approx 3\, 500 $ GeV, $m_{Z'} \approx 4\, 300 $ GeV and $m_{Z'} \approx 5\, 200$ GeV, respectively.
It is essential to mention that in the context of the BLHM, the mass of the  $Z'$ gauge boson depends on the scales of the new model physics, $f$ and $F$. 
From the above, the total cross-section $\sigma^{Z h_0}_{T}\left(\sqrt{s}, f, F \right)$ is sensitive to changes in the free parameters. Contributions from new physics show remarkable effects concerning the SM contribution; the region of most significant appreciation of such effects is for $\sqrt{s}\in [800,10\, 000]$ GeV. 
On the other hand, it is important to mention that Fig.~\ref{sT-fF} has been generated by setting $\tan\, \beta=3$. However, to study the possible dependence of the total cross-section on $\tan\, \beta$, we consider another allowed parameter point, $\tan\, \beta=6$. For this choice, we find that the generated curves are similar to those provided in Fig.~\ref{sT-fF}. Thus, the dependence of the results on $\tan\, \beta$ is null.

\begin{figure}[H]
        \centering
        \includegraphics[width=0.8\textwidth]{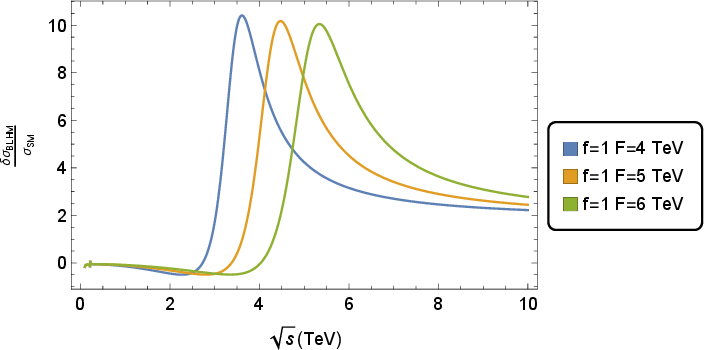}
        \caption{Relative correction $\frac{\delta\sigma_{BLHM}}{\sigma_{SM}}$ as a function of the center-of-mass energy $\sqrt{s}$.}
        \label{Rdeviation-fig}
\end{figure}

The BLHM can generate corrections to the production cross-section for the process  $\mu^{+}\mu^{-} \to Zh_0$  via modification of the tree-level $Z Z h_0$ coupling, as well as by the new interaction vertex  $Z' Z h_0$. 
The values of the relative corrections are calculated from  Eq. (\ref{relative}). 
 In Fig.~\ref{Rdeviation-fig}, we show the relative corrections  $\frac{\delta\sigma_{BLHM}}{\sigma_{SM}}$ of the Higgs-strahlung process  $\mu^{+}\mu^{-} \to Zh_0$ as a function of $\sqrt{s}$ for  $f=1000$ GeV and $F=4 000$ GeV, $f=1 000$ GeV and $F=5000$ GeV, and $f=1 000$ GeV and $F=6 000$ GeV.
 In this figure, the absolute value of the relative correction increases for smaller values of the energy scale $F$ and decouples at high scales of the $\sqrt{s}$ parameter.
The values of $|\frac{\delta\sigma_{BLH}}{\sigma_{SM}}|$ are in the ranges of $0\, \%-10\, \%$ in most of the parameter
space. Our numerical results show that for reasonable values of the free parameters of the BLHM, $\sqrt{s}$,  $f$ and $F$, can generate significant contributions to the total cross-section of the Higgs-strahlung process $\mu^{+}\mu^{-}  \to Zh_0$ concerning their value in the context of the SM.

%\begin{figure}
%        \centering
%        \includegraphics[width=0.7\textwidth]{Fig-9.eps}
%        \caption{Correlation between $F$ and $f$. The contours are for $\sigma_{Tot}=12, 11.5, 11, 10.5, 10\hspace{0.8mm}\rm fb$.}
%        \label{fig:SeS}
%\end{figure}
%

We also discuss the production of $Z h_0$ at the future muon collider, assuming design luminosities ${\cal L}=2, 4, 6, 10, 30\hspace{0.8mm}{\rm ab^{-1}}$ and center-of-mass energies, $\sqrt{s}= 3, 4,  6,  10, 30$ 
% $\sqrt{s}= 3, 4, 5, 6, 7, 10, 30$
TeV~\cite{MuonCollider:2022nsa,MuonCollider:2022xlm,
AlAli:2021let}.  In Tables~\ref{productionzh14}  and~\ref{productionzh16} for $f=1 000$ GeV and $F=4 000$ GeV, and $f=1 000$ GeV and $F=6000$ GeV, respectively, we present an event estimate of the production associated to $Z h_0$.
According to the numerical results, around the resonance of the  $Z'$ gauge boson, the number of $Z h_0$ events reaches high values. In general, the possibility of being observed in the process $\mu^{+} \mu^{-} \to (Z, Z') \to Zh_0$ is quite promising at the future muon collider.

\begin{table}[H]
\caption{The total production of $Zh_0$ at the future muon collider in the context of the BLHM when $\tan \, \beta=3$  with $f=1 000\ \text{GeV}$ and $\ F=4000\ \text{GeV}$ ($m_{Z'}= 3 \, 500$ GeV).
\label{productionzh14}}
    \centering
    \begin{tabular}{|c|c|c|c|c|c|}
    \hline
     \multicolumn{6}{|c|}{ $\tan\, \beta=3$} \\
    \hline
    \multicolumn{6}{|c|}{ $f=1 000$ GeV, $F=4 000$ GeV} \\
    \hline
         $\sqrt{s}$ TeV &$\mathcal{L} = 2$ $ab^{-1}$ & $\mathcal{L} = 4$ $ab^{-1}$  & $\mathcal{L} = 6$ $ab^{-1}$ & $\mathcal{L} = 10$ $ab^{-1}$ & $\mathcal{L} = 30$ $ab^{-1}$  \\
         \hline
         3 & 7 405 & 14 810 & 22 215 & 37 026 & 111 079  \\
         \hline
         4 & 13 882 & 27 764 & 41 646 & 69 410 & 208 231  \\
         \hline
        % 5 & 5 123 & 10 247 & 15 371 & 25 618 & 76 855  \\
%         \hline
         6 & 2 822 & 5 645 & 8 468 & 14 113 & 42 341  \\
         \hline
      %    7 & 1 847 & 3 694 & 5 541 & 9 235 & 27 705  \\
%         \hline
          10 & 787 & 1 575 & 2 363 & 3 938 & 11 861  \\
         \hline
          30 & 79 & 158 & 237 & 396 & 1 188  \\
         \hline

    \end{tabular}

\end{table}

\begin{table}[H]
\caption{The total production of $Zh_0$ at the future muon collider in the context of the BLHM when $\tan \, \beta=3$  with $f=1 000\ \text{GeV}$ and $\ F=6 000\ \text{GeV}$ ($m_{Z'}=5\ 200$ GeV).
\label{productionzh16}}
    \centering
    \begin{tabular}{|c|c|c|c|c|c|}
    \hline
     \multicolumn{6}{|c|}{ $\tan\, \beta=3$} \\
    \hline
    \multicolumn{6}{|c|}{ $f=1 000$ GeV, $F=6 000$ GeV} \\
    \hline
         $\sqrt{s}$ TeV &$\mathcal{L} = 2$ $ab^{-1}$ & $\mathcal{L} = 4$ $ab^{-1}$  & $\mathcal{L} = 6$ $ab^{-1}$ & $\mathcal{L} = 10$ $ab^{-1}$ & $\mathcal{L} = 30$ $ab^{-1}$  \\
         \hline
         3 & 1 501 & 3 003 & 4 505 & 7 509 & 22 529  \\
         \hline
         4 & 1 447 & 2 895 & 4 343 & 7 239 & 21 717  \\
         \hline
       %  5 & 8 903 & 17 806 & 26 709 & 44 515 & 133 546  \\
%         \hline
         6 & 5 793 & 11 586 & 17 380 & 28 966 & 86 900  \\
         \hline
         % 7 & 2 871 & 5 742 & 8 613 & 14 356 & 43 068  \\
%         \hline
          10 & 922 & 1 485 & 2 768 & 4 614 & 13 844 \\
         \hline
          30 & 80 & 160 & 240 & 401 & 1 204  \\
         \hline
    \end{tabular}

\end{table}

\subsection{Heavy Higgs boson production $H_0$ in the BLHM}

\begin{figure}[H]
        \centering
        \includegraphics[width=0.8\textwidth]{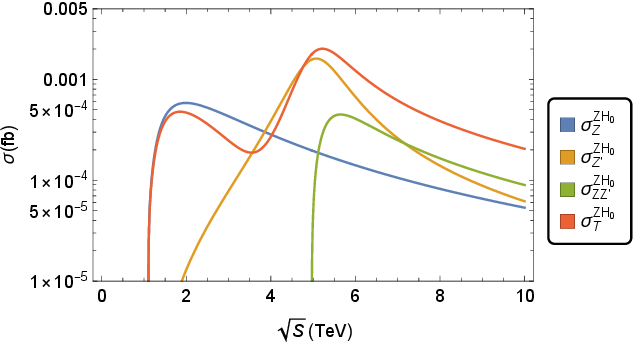}
        \caption{The cross-section of the process $\mu^{+}\mu^{-}\to (Z, Z') \to ZH_0$  as a function of $\sqrt{s}$. 
               The curves are generated for   $f = 1 000$ GeV and $F=6 000$ GeV (for $m_{Z'} = 5\ 200$ GeV), and correspond to  $\sigma^{Z H_0}_Z$ (Eq.~(\ref{ZH0-Z})), $\sigma^{Z H_0}_{Z'}$ (Eq. (\ref{ZH0-Zp})), $\sigma^{Z H_0}_{ZZ'}$ (Eq. (\ref{ZH0-mezcla})), and  $ \sigma^{Z H_0}_{T} $ (Eq.~(\ref{ZH0tot})).}
        \label{S-ZH0}
\end{figure}

In this subsection, we present our results on the production cross-section of the Higgs-strahlung process $\mu^{+}\mu^{-}\to (Z, Z') \to ZH_0$, and analyze the impact of the parameters of the BLHM on this process. 
From Fig.~\ref{strahlung}, we can observe that the total cross-section of the production process $\mu^{+}\mu^{-} \to ZH_0$ receives contributions from the  $Z$ and $Z'$ gauge bosons, and from the interference effects between them.  
In this manner,  cross-sections  $\sigma^{Z H_0}_{Z}\left(\mu^{+} \mu^{-} \to Z H_0 \right)$, $\sigma^{Z H_0}_{Z'}\left(\mu^{+} \mu^{-} \to Z H_0 \right)$ and $\sigma^{Z H_0}_{Z Z'}\left(\mu^{+} \mu^{-} \to Z H_0 \right)$  contribute to the total cross-section $\sigma^{Z H_0}_{T}\left(\mu^{+} \mu^{-} \to Z H_0 \right)$. These production cross-sections depend on the free parameters of the BLHM: $f$, $F$, and $\sqrt{s}$. Thus, in order to analyze the effects of $\sqrt{s}$ on $\sigma^{Z H_0}_{i}\left(\mu^{+} \mu^{-} \to Z H_0  \right)$ with $i=Z,Z',ZZ',T$, we generate the curves in Fig.~\ref{S-ZH0} by setting the other input parameters to $f=1 000$ GeV and $F=6 000$ GeV. For these elections, the new heavy gauge boson $Z'$ obtains a mass of about $5\, 200$ GeV. 
 In this figure, we observe that the curves associated to $\sigma^{Z H_0}_{Z'}$, $\sigma^{Z H_0}_{ZZ'}$, and $ \sigma^{ZH_0}_{T}$ obtain large values around the resonance energy of the $Z'$ boson, specifically, for the values of $\sqrt{s}\approx 5\, 200, 5\, 600, 5\, 200$ GeV generate $\sigma^{ZH_0}_{Z'}\left(\mu^{+} \mu^{-} \to Z H_0  \right)=1.60\times 10^{-3}$ fb, $\sigma^{Z H_0}_{ZZ'}\left(\mu^{+} \mu^{-} \to Z H_0  \right)=4.47\times 10^{-4}$ fb and $\sigma^{Z H_0}_{T}\left(\mu^{+} \mu^{-} \to Z H_0  \right)=2.01\times 10^{-3}$ fb, respectively.
Regarding the cross-section $\sigma^{Z H_0}_{Z}\left(\mu^{+} \mu^{-} \to Z H_0  \right)$, it reaches the maximum when $\sqrt{s}\approx 2\, 000 $ GeV, at this point $\sigma^{Z H_0}_{Z}\left(\mu^{+} \mu^{-} \to Z H_0  \right)=5.84\times 10^{-4}$ fb. 
As discussed in Subsection~\ref{ph0}, in specific regions, the total cross-section of the process $\mu^+\mu^- \to (Z,Z') \to ZH_0$ is smaller than the individual contributions. This is primarily due to negative interference between the channels $\mu^+\mu^- \to Z \to ZH_0$ and $\mu^+\mu^- \to Z' \to ZH_0$. This effect is further influenced by the effective couplings $g_{ZZ'H_0}$ and $g_{ZZH_0}$ (see Appendix~\ref{rulesF}).

%In Fig.~\ref{Stot-ZH0}, we also present the production cross-section of the process $\mu^{+}\mu^{-} \to ZH_0$  as a function of $\sqrt{s}$ while the energy scales, $f$ and $F$, take on certain fixed values:  $f=1 000$ GeV and $F=4 000$ GeV, $f=1 000$ GeV and $F=5 000$ GeV, and $f=1 000$ GeV and $F=6 000$ GeV.
%From this figure, a slight increase of the total cross-section $\sigma^{Z H_0}_{T}\left( \mu^{+}\mu^{-} \to ZH_0 \right)$ is observed for small values of the $F$ scale while for large values of the center-of-mass energy $\sqrt{s}$, $\sigma^{Z H_0}_{T}\left( \mu^{+}\mu^{-} \to ZH_0 \right)$ becomes smaller. For the plotted curves, resonant effects dominate, i.e., the maximum peaks of each curve are reached just at the resonance of the $Z'$ gauge boson: $\sigma^{Z H_0}_{T}\left(\sqrt{s},1\, 000\ \text{GeV}, 4\, 000\ \text{GeV} \right)=3.93 \times 10^{-3}$ fb, $\sigma^{Z H_0}_{T}\left(\sqrt{s},1\, 000\ \text{GeV}, 5\, 000\ \text{GeV} \right)=2.72 \times 10^{-3}$ fb, and $\sigma^{Z H_0}_{T}\left(\sqrt{s},1\, 000\ \text{GeV}, 6\, 000\ \text{GeV} \right)=2.01 \times 10^{-3}$ fb for the corresponding energies, $\sqrt{s} \approx 3\, 500,\, 4\, 300,\, 5\, 200$ GeV.
%In the BLHM scenario, the measure of the fine-tuning depends only on the $f$ scale, as discussed above, for values of $f \in [1\, 000,2\, 000]$ GeV, the absence of the fine-tuning prevails in the model, and for this reason, our plots have been generated for $f=1\, 000$ GeV while allowing the other free parameters to vary. 

\begin{figure}[H]
        \centering
        \includegraphics[width=0.9\textwidth]{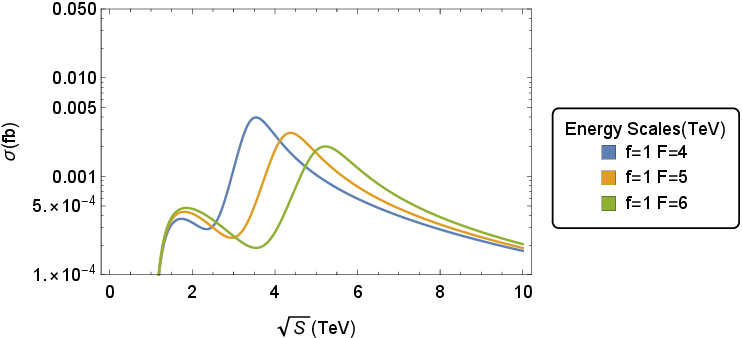}
        \caption{The total cross-section of the process $\mu^{+}\mu^{-}\to (Z, Z') \to ZH_0$  as a function of $\sqrt{s}$.
              The curves are generated for $f=1 000$ GeV and $F=4 000$ GeV, $f=1 000$ GeV and $F=5 000$ GeV, and $f=1 000$ GeV and $F=6 000$ GeV. In each case, $\tan\, \beta=3$ and $ m_{H_{0}} \approx 1015\  \text{GeV} $ have been considered.}
        \label{Stot-ZH0}
\end{figure}

\begin{figure}[H]
        \centering
        \includegraphics[width=0.9\textwidth]{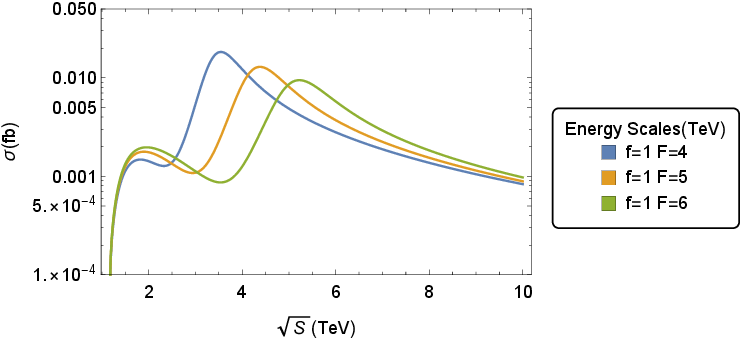}
        \caption{The total cross-section of the process $\mu^{+}\mu^{-}\to (Z, Z') \to ZH_0$  as a function of $\sqrt{s}$.
              The curves are generated for $f=1 000$ GeV and $F=4 000$ GeV, $f=1 000$ GeV and $F=5 000$ GeV, and $f=1 000$ GeV and $F=6 000$ GeV. In each case, $\tan\, \beta=6$ and $ m_{H_{0}} \approx 1076\  \text{GeV} $ have been considered.}
        \label{Stot-ZH0-tan6}
\end{figure}

In Figs.~\ref{Stot-ZH0} and~\ref{Stot-ZH0-tan6}, we also analyze the dependence of the production cross-section of the process $\mu^{+}\mu^{-} \to ZH_0$ on $\tan\, \beta$. These figures have been generated considering the parameter points, $\tan\, \beta=3$ and $\tan\, \beta=6$, respectively.   For these cases, the total cross-section is 
a function of $\sqrt{s}$ while the energy scales, $f$ and $F$, take on certain fixed values:  $f=1 000$ GeV and $F=4 000$ GeV, $f=1 000$ GeV and $F=5 000$ GeV, and $f=1 000$ GeV and $F=6 000$ GeV.
Regarding Fig.~\ref{Stot-ZH0}, a slight increase of the total cross-section $\sigma^{Z H_0}_{T}\left( \mu^{+}\mu^{-} \to ZH_0 \right)$ is observed for small values of the $F$ scale while for large values of the center-of-mass energy $\sqrt{s}$, $\sigma^{Z H_0}_{T}\left( \mu^{+}\mu^{-} \to ZH_0 \right)$ becomes smaller. For the plotted curves, resonant effects dominate, i.e., the maximum peaks of each curve are reached just at the resonance of the $Z'$ gauge boson: $\sigma^{Z H_0}_{T}\left(\sqrt{s},1\, 000\ \text{GeV}, 4\, 000\ \text{GeV} \right)=3.93 \times 10^{-3}$ fb, $\sigma^{Z H_0}_{T}\left(\sqrt{s},1\, 000\ \text{GeV}, 5\, 000\ \text{GeV} \right)=2.72 \times 10^{-3}$ fb, and $\sigma^{Z H_0}_{T}\left(\sqrt{s},1\, 000\ \text{GeV}, 6\, 000\ \text{GeV} \right)=2.01 \times 10^{-3}$ fb for the corresponding energies, $\sqrt{s} \approx 3\, 500,\, 4\, 300,\, 5\, 200$ GeV.
With respect to Fig.~\ref{Stot-ZH0-tan6}, obtained for $\tan\, \beta=6$, we can appreciate that the height of the maximum peaks of each curve is reached, again, just at the resonance of the $Z'$ gauge boson: $\sigma^{Z H_0}_{T}\left(\sqrt{s},1\, 000\ \text{GeV}, 4\, 000\ \text{GeV} \right)=1.82 \times 10^{-2}$ fb, $\sigma^{Z H_0}_{T}\left(\sqrt{s},1\, 000\ \text{GeV}, 5\, 000\ \text{GeV} \right)=1.29 \times 10^{-2}$ fb, and $\sigma^{Z H_0}_{T}\left(\sqrt{s},1\, 000\ \text{GeV}, 6\, 000\ \text{GeV} \right)=9.46 \times 10^{-3}$ fb when $\sqrt{s} \approx 3\, 500,\, 4\, 300,\, 5\, 200$ GeV, respectively.  For this case, the height of the maximum peaks of the plotted curves is higher than the maximum peaks reached when $\tan\, \beta=3$.
This occurs because $m_{H_0}$, a parameter involved in the calculation of the total cross-section $\sigma^{Z H_0}_{T}\left( \mu^{+}\mu^{-} \to ZH_0 \right)$, has a dependence on the parameter $\beta$ (see Eq.~(\ref{massH0})).
In the BLHM scenario, the fine-tuning measurement depends on the $f$ energy scale. As discussed above, for values of $f$ close to 1000 GeV, the absence of the fine-tuning prevails in the model. For this reason, our plots have been generated for $f=1\, 000$ GeV while allowing the other free parameters to vary.

As part of our study, and as an indicator of the possible number of $Z H_0$ events to be produced in a future muon collider, we consider again the center-of-mass energies $\sqrt{s}= 3, 4, 6, 10, 30$ 
% $\sqrt{s}= 3, 4, 5, 6, 7, 10, 30$
TeV and the integrated luminosities ${\cal L}=2, 4, 6, 10, 30$ $\rm ab^{-1}$~\cite{MuonCollider:2022nsa,MuonCollider:2022xlm,AlAli:2021let}. 
For the two scenarios discussed above, $\tan \, \beta=3$ and $\tan \, \beta=6$,
 in Tables~\ref{TableZH014}-\ref{TableZH016-tan6} we list the number of $Z H_0$ events arising when $f=1 000$ GeV and $F=4 000$ GeV, and $f=1 000$ GeV and $F=6 000$ GeV.  
 According to our numerical data, the possibility of performing measurements for the $Z'$ gauge boson and the heavy Higgs boson $H_0$ at the future high-energy muon collider is modest.  
 For these cases of interest, resonant effects dominate over non-resonant effects. Thus, the total cross-section $\sigma^{Z H_0}_{T}\left( \mu^{+}\mu^{-} \to ZH_0 \right)$ reaches its maximum value at the resonance of the heavy gauge boson $Z'$.

\begin{table}[H]
\caption{The total production of $ZH_0$ at the future muon collider in the context of the BLHM when $\tan \, \beta=3$ with $f=1000\ \text{GeV}$ and $\ F=4 000\ \text{GeV}$ ($m_{Z'}=3\, 500$ GeV).
\label{TableZH014}}
    \centering
    \begin{tabular}{|c|c|c|c|c|c|}
    \hline
    \multicolumn{6}{|c|}{ $\tan\, \beta=3$} \\
    \hline
    \multicolumn{6}{|c|}{ $f=1 000$ GeV, $F=4 000$ GeV} \\
    \hline
         $\sqrt{s}$ TeV &$\mathcal{L} = 2$ $ab^{-1}$ & $\mathcal{L} = 4$ $ab^{-1}$  & $\mathcal{L} = 6$ $ab^{-1}$ & $\mathcal{L} = 10$ $ab^{-1}$ & $\mathcal{L} = 30$ $ab^{-1}$  \\
         \hline
         3 & 2  & 4 & 7 & 11 & 35  \\
         \hline
         4 & 5 & 10 & 15 & 26 & 78  \\
         \hline
         %5 & 2 & 4 & 6 & 10 & 31  \\
%         \hline
         6 & 1 & 2 & 3 & 5 & 17  \\
         \hline
        %  7 & 1 & 1 & 2 & 4 & 12  \\
%         \hline
          10 & 1 & 1 & 1 & 2 & 5 \\
         \hline
          30 & 1 & 1 & 1 & 1 & 1   \\
         \hline

    \end{tabular}

\end{table}

\begin{table}[H]
\caption{The total production of $ZH_0$ at the future muon collider in the context of the BLHM when $\tan \, \beta=3$ with $f=1 000\ \text{GeV}$ and $\ F=6 000\ \text{GeV}$ ($m_{Z'}=5\, 200$ GeV).
\label{TableZH016}}
    \centering
    \begin{tabular}{|c|c|c|c|c|c|}
    \hline
     \multicolumn{6}{|c|}{ $\tan\, \beta=3$} \\
    \hline
    \multicolumn{6}{|c|}{ $f=1 000$ GeV, $F=6 000$ GeV} \\
    \hline
         $\sqrt{s}$ TeV &$\mathcal{L} = 2$ $ab^{-1}$ & $\mathcal{L} = 4$ $ab^{-1}$  & $\mathcal{L} = 6$ $ab^{-1}$ & $\mathcal{L} = 10$ $ab^{-1}$ & $\mathcal{L} = 30$ $ab^{-1}$  \\
         \hline
         3 & 1  & 1 & 1 & 2 & 7  \\
         \hline
         4 & 1 & 1 & 1 & 2 & 8  \\
         \hline
         %5 & 3 & 7 & 10 & 18 & 54  \\
%         \hline
         6 & 2 & 4 & 7 & 12 & 36  \\
         \hline
         % 7 & 1 & 2 & 3 & 6 & 18  \\
%         \hline
          10 & 1 & 1 & 2 & 2 & 6 \\
         \hline
          30 & 1 & 1 & 1 & 1 & 1   \\
         \hline

    \end{tabular}

\end{table}

\begin{table}[H]
\caption{The total production of $ZH_0$ at the future muon collider in the context of the BLHM when $\tan \, \beta=6$ with $f=1000\ \text{GeV}$ and $\ F=4 000\ \text{GeV}$ ($m_{Z'}=3\, 500$ GeV).
\label{TableZH014-tan6}}
    \centering
    \begin{tabular}{|c|c|c|c|c|c|}
    \hline
    \multicolumn{6}{|c|}{ $\tan\, \beta=6$} \\
    \hline
    \multicolumn{6}{|c|}{ $f=1 000$ GeV, $F=4 000$ GeV} \\
    \hline
         $\sqrt{s}$ TeV &$\mathcal{L} = 2$ $ab^{-1}$ & $\mathcal{L} = 4$ $ab^{-1}$  & $\mathcal{L} = 6$ $ab^{-1}$ & $\mathcal{L} = 10$ $ab^{-1}$ & $\mathcal{L} = 30$ $ab^{-1}$  \\
         \hline
         3 & 10  & 21 & 32 & 53 & 161  \\
         \hline
         4 & 24 & 48 & 72 & 121 & 364  \\
         \hline
         6 & 5 & 11 & 16 & 27 & 83  \\
         \hline
          10 & 1 & 3 & 4 & 8 & 24 \\
         \hline
          30 & 1 & 1 & 1 & 1 & 2   \\
         \hline
    \end{tabular}

\end{table}

\begin{table}[H]
\caption{The total production of $ZH_0$ at the future muon collider in the context of the BLHM when $\tan \, \beta=6$ with $f=1 000\ \text{GeV}$ and $\ F=6 000\ \text{GeV}$ ($m_{Z'}=5\, 200$ GeV).
\label{TableZH016-tan6}}
    \centering
    \begin{tabular}{|c|c|c|c|c|c|}
    \hline
     \multicolumn{6}{|c|}{ $\tan\, \beta=6$} \\
    \hline
    \multicolumn{6}{|c|}{ $f=1 000$ GeV, $F=6 000$ GeV} \\
    \hline
         $\sqrt{s}$ TeV &$\mathcal{L} = 2$ $ab^{-1}$ & $\mathcal{L} = 4$ $ab^{-1}$  & $\mathcal{L} = 6$ $ab^{-1}$ & $\mathcal{L} = 10$ $ab^{-1}$ & $\mathcal{L} = 30$ $ab^{-1}$  \\
         \hline
         3 & 2  & 4 & 6 & 11 & 33  \\
         \hline
         4 & 2 & 5 & 7 & 12 & 37  \\
         \hline
         6 & 11 & 22 & 34 & 57 & 172  \\
         \hline
          10 & 1 & 3 & 5 & 9 & 29 \\
         \hline
          30 & 1 & 1 & 1 & 1 & 2   \\
         \hline
    \end{tabular}
\end{table}

We also include an analysis of the final signal and SM background for each  of the processes studied $\mu^{+}\mu^{-} \to Zh_0$ and $\mu^{+}\mu^{-} \to ZH_0$. Specifically, we provide the total cross-section and the corresponding  number of events when considering the most important SM background of the processes $\mu^{+}\mu^{-} \to Zh_0$ and $\mu^{+}\mu^{-} \to ZH_0$. The background of the processes studied in our article are described below.

\subsection{Process $\mu^{+}\mu^{-} \to Zh_0$: Final signal and SM background }

Because the Higgs boson’s decay rate to $b\bar b$ is greater than the decay rate to other quarks and leptons, the $b\bar b$ decay mode of Higgs
$(h_0 \to b\bar b)$ is considered. Since the cross-sections of the background processes corresponding to the leptonic decays of the $Z$ boson are
less than the background cross-sections corresponding to the other decays, the leptonic decays of the $Z$ boson in the Higgs-strahlung process
are taken into account. The signal process of interest is
 $\mu^{+}\mu^{-} \to Zh_0 \to l^+l^- b\bar b$  ($l^{-}=e^{-}, \mu^{-}$)  with the background processes $\mu^{+}\mu^{-} \to ZZ, Z\gamma, \gamma\gamma$.
%  $\mu^{+}\mu^{-} \to W^+W^-Z,  W^+W^-\gamma$.
  
For the analysis, we start from the narrow-width approximation which is a useful way to simplify the calculation of complicated processes.
Therefore, we apply this method to determine the total cross-section of the $\sigma \bigl(\mu^{+}\mu^{-} \to Zh_0 \to l^+l^- b\bar b)$
signal,

\begin{eqnarray}
\sigma \bigl(\mu^{+}\mu^{-} \to Zh_0 \to l^+l^- b\bar b)&\simeq&\sigma \bigl(\mu^+\mu^- \to Zh_0\bigr)\times \text{Br}(Z \to l^+ l^-)\nonumber \\
&\times& \text{Br}(h_0 \to b \bar b).
\end{eqnarray}

\noindent An SM Higgs boson of mass 125 GeV has a $60\%$ branching ratio to the final state $b\bar b$, $\text{Br}(h_0 \to b\bar b)=60\%$~\cite{ParticleDataGroup:2024cfk}. Therefore, the representative cross-sections for the dominant background of the $Zh_0$ final state at the future muon collider shown in Table~\ref{h00}.

\begin{table}[H]
\caption{Representative cross-sections for the background of the $Zh_0$ final state at the future muon collider for the
center-of-mass energies of $\sqrt{s}=3, 4, 6, 10, 30\hspace{0.8mm}$ TeV.}
\label{h00}
\begin{center}
 \begin{tabular}{|c|c|c|c|c|c|}
\hline\hline
\multicolumn{6}{|c|}{Background cross-section [fb]}\\
 \hline\hline
% \multicolumn{5}{|c|}{Number of expected events at the future muon collider}\\
%\hline
\cline{1-6} {\rm Process}   &  $\sqrt{s}=3$ TeV &  $\sqrt{s}=4$ TeV  & $\sqrt{s}=6$ TeV   & $\sqrt{s}=10$ TeV  & $\sqrt{s}=30$ TeV   \\
\hline
       $\mu^+\mu^- \to ZZ$  &  10.65            &  14.20             & 21.31               & 35.52             & 106.54                \\

\hline\hline
\end{tabular}
\end{center}
\end{table}

In Table~\ref{bzh0}, we present the number of events for $\mu^{+}\mu^{-} \to Zh_0 \to l^+l^- b\bar b$ signal and consider the dominant background. 
Furthermore, the $l^+l^- b\bar b$ process can be mimicked by diboson production at the future muon collider with sizable rates (see Table~\ref{h00}).
In this way, more complete results on the total production in the number of events are presented below in Table~\ref{bzh0}.

\begin{table}[H]
\caption{The total production of $\mu^{+}\mu^{-} \to (Z, Z') \to Zh_0 \to l^+l^- b\bar b$ and dominant background $ZZ$ at the future muon collider
in the context of the BLHM when $\tan \, \beta=3$  with $f=1 000\ \text{GeV}$ and $\ F=4000\ \text{GeV}$ ($m_{Z'}= 3 \, 500$ GeV).
}
\label{bzh0}
    \centering
    \begin{tabular}{|c|c|c|c|c|c|}
    \hline
     \multicolumn{6}{|c|}{ $\tan\, \beta=3$} \\
    \hline
    \multicolumn{6}{|c|}{ $f=1 000$ GeV, $F=4 000$ GeV} \\
    \hline
         $\sqrt{s}$ TeV &$\mathcal{L} = 2$ $ab^{-1}$ & $\mathcal{L} = 4$ $ab^{-1}$  & $\mathcal{L} = 6$ $ab^{-1}$ & $\mathcal{L} = 10$ $ab^{-1}$ & $\mathcal{L} = 30$ $ab^{-1}$  \\
         \hline
         3 & 1 119 & 2 238 & 3 358 & 5 597 & 16 792  \\
         \hline
         4 & 1 648 & 3 297 & 4 946 & 8 244 & 24 738  \\
         \hline
        % 5 & 5 123 & 10 247 & 15 371 & 25 618 & 76 855  \\
%         \hline
         6 & 1 772 & 3 544 & 5 316 & 8 861 & 26 583  \\
         \hline
      %    7 & 1 847 & 3 694 & 5 541 & 9 235 & 27 705  \\
%         \hline
          10 & 2 801 & 5 602 & 8 403 & 14 006 & 42 020  \\
         \hline
          30 & 8 313 & 16 626 & 24 939 & 41 566 & 124 698  \\
         \hline
    \end{tabular}
\end{table}

\subsection{Process $\mu^{+}\mu^{-} \to ZH_0 \to l^+l^- W^+W^-$: Final signal and SM background }

\begin{figure}[H]
\center
{\includegraphics[width=12.10cm]{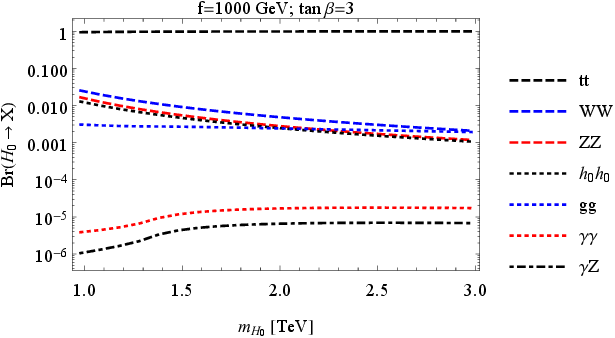}}
\caption{\label{brH0} The branching ratio for the $H_0\to X$ process  as a function of the $m_{H_0}$ parameter where  $X \equiv $ $t\bar{t}$,  $WW$, $ZZ$, $h_0h_0$, $gg$,  $\gamma\gamma$, $\gamma Z$.
}
\end{figure}

 In the case of the Higgs-strahlung process $\mu^{+}\mu^{-} \to ZH_0$  (with the heavy Higgs boson $H_0$ predicted by the BLHM). For the final state with $W^\pm$ bosons, the signal process is $\mu^{+}\mu^{-} \to ZH_0 \to l^+l^- W^+W^-$ with the background processes $\mu^{+}\mu^{-} \to W^{+}W^{-}Z, W^{+}W^{-} \gamma$. In this case,  it is necessary to evaluate the branching ratio for the $H_0 \to X$ process  ($X \equiv t\bar t, WW, ZZ, h_0h_0, gg,\gamma\gamma, \gamma Z$) to realize the analysis of the final signal and SM background. Therefore, in Fig.~\ref{brH0}, we presented the $\text{Br}(H_0\to X)$ vs. $m_{H_0}$ when  $f=1000$ GeV and $\tan\beta=3$. From this figure, it is clear that the subdominant channel of the heavy
Higgs boson decay is to a pair of $W^\pm$-bosons, followed by pairs of $Z$-bosons, etc..

\noindent With these elements, we apply the narrow-width approximation to determine the total cross-section of the $\mu^{+}\mu^{-} \to ZH_0 \to l^+l^- W^+W^-$ signal,
 
\begin{eqnarray}
\sigma \bigl(\mu^{+}\mu^{-} \to ZH_0 \to l^+l^- W^+W^-)&\simeq&\sigma \bigl(\mu^+\mu^- \to ZH_0\bigr)\times \text{Br}(Z \to l^+ l^-)\nonumber \\
&\times& \text{Br}(H_0 \to W^+W^-).
\end{eqnarray}

\noindent  As an illustration, we only consider the most important background for the process $\mu^{+}\mu^{-} \to ZH_0 \to l^+l^- W^+W^-$. Remember (see Fig.~\ref{brH0}) that the heavy 
Higgs boson subdominant decay is to pairs of $W^\pm$-bosons. Therefore, the cross-sections of the most important background of the $\mu^{+}\mu^{-} \to ZH_0 \to l^+l^- W^+W^-$ signal is shown in Table~\ref{H00}.

\begin{table}[H]
\caption{Representative cross-sections for the background of the $ZH_0$ final state at the future muon collider for the
center-of-mass energies of $\sqrt{s}=3, 4, 6, 10, 30\hspace{0.8mm}$ TeV.}
\label{H00}
\begin{center}
 \begin{tabular}{|c|c|c|c|c|c|}
\hline\hline
\multicolumn{6}{|c|}{Background cross-section [fb]}\\
 \hline\hline
% \multicolumn{5}{|c|}{Number of expected events at the future muon collider}\\
%\hline
\cline{1-6} {\rm Process}   &  $\sqrt{s}=3$ TeV &  $\sqrt{s}=4$ TeV  & $\sqrt{s}=6$ TeV   & $\sqrt{s}=10$ TeV  & $\sqrt{s}=30$ TeV   \\
\hline
       $\mu^+\mu^- \to WWZ$ &  31               &  41.3              & 62                 & 103.3              & 310                \\

\hline\hline
\end{tabular}
\end{center}
\end{table}

The total production in the number of events is presented below in Table~\ref{bzH0} for $\sqrt{s}=3 000, 10 000$ GeV and $\mathcal{L} = 3, 4, 6, 10, 30$ ${\rm ab}^{-1}$. 
In addition, we consider the Leptonic, Semi-leptonic and Hadronic channels of the $W^\pm$ for the signal. Thus, we assume that the branching ratios for $W^\pm$ decays are: $\text{Br}(W^\pm \to qq') = 0.454$ for hadronic decays, $\text{Br}(W^\pm \to qq'; W^\pm \to l\nu_{e, \mu}) = 0.143$ for semi-leptonic decays and $\text{Br}(W^\pm \to l\nu_{e, \mu}) = 0.045$ for light leptonic decays~\cite{ParticleDataGroup:2024cfk}.

\begin{table}[H]
\caption{The total production of $\mu^{+}\mu^{-} \to (Z, Z') \to ZH_0 \to l^+l^- W^+W^-$ and dominant background $WWZ$ at the future muon collider
in the context of the BLHM when $\tan \, \beta=3$  with $f=1 000\ \text{GeV}$ and $\ F=4000\ \text{GeV}$ ($m_{Z'}= 3 \, 500$ GeV). The Leptonic, Semi-leptonic, and Hadronic channels of the $W^\pm W^\mp$ in the final state are considered.}
\label{bzH0}
\begin{center}
\begin{tabular}{|c|c|c|c|}
\hline\hline
 \multicolumn{4}{|c|}{ $\tan\, \beta=3$} \\
\hline
  & \multicolumn{3}{|c|}{ $\sqrt{s}=3 000$ GeV,  $f=1 000$ GeV, $F=4 000$ GeV } \\
\cline{2-4}
 ${\cal L} \, (\rm ab^{-1})$  & Leptonic channel & Semi-leptonic channel& Hadronic channel \\
\hline\hline
\cline{1-4}
             3        &    167     &     532       &   1 688        \\
 \hline
             4        &    335     &    1 064      &   3 377        \\
 \hline
             6        &    502     &    1 596      &   5 066        \\
 \hline
             10       &    837     &    2 660      &   8 442        \\
 \hline
             30       &    2 511   &    7 979      &  25 333        \\
 \hline
 \hline
             & \multicolumn{3}{|c|}{ $\sqrt{s} = 10 000$  GeV, $f=1 000$ GeV, $F=4 000$ GeV } \\
\cline{2-4}
 ${\cal L} \, (\rm ab^{-1})$  & Leptonic channel & Semi-leptonic channel& Hadronic channel \\
\hline\hline
\cline{1-4}
             3        &    558     &     1 772       &    5 628        \\
 \hline
             4        &   1 115    &     3 545       &   11 255        \\
 \hline
             6        &   1 673    &     5 318       &   16 883        \\
 \hline
             10       &   2 789    &     8 863       &   28 139        \\
 \hline
             30       &   8 367    &    26 589       &   84 417        \\
 \hline
 \hline 
%           & \multicolumn{3}{|c|}{ $\sqrt{s} = 7$  TeV, \hspace{5mm} $M_{Z'}=7$ TeV, \hspace{5mm} $g'_1=0.93$ } \\
%\cline{2-4}
% ${\cal L} \, (\rm ab^{-1})$  & Leptonic channel & Semi-leptonic channel& Hadronic channel \\
%\hline\hline
%\cline{1-4}
%             1        &    3 257     &     10 348       &   32 955        \\
% \hline
%             2        &    5 513     &     20 597       &   65 710        \\
% \hline
%             3        &    9 770     &     31 046       &   98 562        \\
% \hline
%             4        &   13 026     &     41 394       &  131 420        \\
% \hline
%             10       &   32 566     &    103 486       &  329 551        \\
% \hline
% \hline
\end{tabular}
\end{center}
\end{table}

\section{Conclusions} \label{conclusions}

In this article, we have studied the  $Z'$ boson of the BLHM as a portal to signatures of Higgs bosons $h_0$ and $H_0$
through the Higgs-strahlung production processes $\mu^+\mu^- \to (Z,Z') \to Zh_0, ZH_0$, including both the resonant and non-resonant effects. 
The new $Z'$ boson is a hypothetical massive particle of spin 1 that is also predicted in other extensions of the SM and has been the subject of extensive phenomenological studies in recent years~\cite{Leike:1998wr}.  Experimentally, the $Z'$ gauge boson will be searched at the LHC~\cite{ParticleDataGroup:2024cfk}.
In the context of the BLHM, Higgs-strahlung productions $\mu^+\mu^- \to Zh_0, ZH_0$ are essential processes to study tree-level interactions: $Z'Z h_0$, $Z'Z H_0$, $Z Z h_0$ and $Z Z H_0$.  At the same time, the mentioned processes are helpful to test the consistency of the current parameter space of the BLHM.
For example, through the Higgs-strahlung process $\mu^+\mu^- \to Z \to Zh_0$, we have found that it reproduces very well the SM predictions when the new physics scales, $f$ and $F$, take large values (see Fig.~\ref{sigma-zh0}). For this case, the effective couplings $g^{\text{BLHM}}_{ZZh_0} \approx g^{\text{SM}}_{ZZh_0}$.

As for the Higgs-strahlung process $\mu^+\mu^- \to (Z,Z') \to Zh_0 $, for its study we consider the BLHM contributions generated through the $Z'Z h_0$ and $ZZ h_0$ couplings. We find that the relative correction of the total cross-section $\sigma^{Z h_0}_{T} \left(\mu^+\mu^- \to Zh_0 \right)$ from its SM prediction can vary from 0 to 10$\%$, which arises mainly from the modifications of the $Z Z h_0$ coupling, and also from the contribution induced through the $Z'Z h_0$ interaction vertex. Our numerical data show that for reasonable values of the free parameters of the BLHM it can generate significant contributions to $\sigma^{Z h_0}_{T} \left(\mu^+\mu^- \to Zh_0 \right)$. In most of the parameter space, the relative corrections values are positive and decoupled at high scales from the $\sqrt{s}$ parameter.

In the BLHM, we explore the phenomenological implications of the production cross-section of the processes $\mu^+\mu^-  \to Zh_0, ZH_0$. As a result of our analysis, we find that the cross-sections $\sigma^{Z h_0}_{T} \left(\mu^+\mu^- \to Zh_0 \right)$ and $\sigma^{Z H_0}_{T} \left(\mu^+\mu^- \to Z H_0 \right)$ reach large values at the resonance of the heavy gauge boson $Z'$,  when $\sqrt{s}=m_{Z'}$.  $\sigma^{Z h_0}_{T} \left(\mu^+\mu^- \to Zh_0 \right)$ and $\sigma^{Z H_0}_{T} \left(\mu^+\mu^- \to Z H_0 \right)$ are also sensitive to variations in the $F$ parameter, 
and the height of the resonance peaks for the $Z'$ boson changes depending on the $F$ scale values.
Thus, the cross-sections $\sigma^{Z h_0}_{T} \left(\mu^+\mu^- \to Zh_0 \right)$ and $\sigma^{Z H_0}_{T} \left(\mu^+\mu^- \to Z H_0 \right)$  obtain large values when $F$ obtains small values.  Another input parameter involved in our cross-section calculations is the $f$ scale, which is set to 1000 GeV to ensure the absence of fine-tuning in our phenomenological predictions.

To estimate the production of Higgs bosons $h_0$ and $H_0$ at the future muon collider, we use the energies and design luminosities of the muon collider with the center-of-mass energies of  $\sqrt{s}=3, 4,  6, 10, 30$
% $\sqrt{s}= 3, 4, 5, 6, 7, 10, 30$
 TeV and integrated luminosities of  ${\cal L}=2, 4, 6, 10, 30$ $\rm ab^{-1}$~\cite{MuonCollider:2022nsa,MuonCollider:2022xlm,AlAli:2021let}.
We can observe from  Tables~\ref{productionzh14}-\ref{TableZH016-tan6} that the total number of expected events for $Z h_0$ and $ZH_0$ at a future muon collider increase just at the resonance energy of the $Z'$ boson. 
  Our results show a very optimistic scenario for producing Higgs bosons $h_0$ and bosons $Z$ in the future muon experiment.   
Regarding the production of  Higgs bosons $H_0$ and bosons $Z$, these show a more conservative scenario.

It is worth mentioning that in our study we have incorporated the final signal and SM background corresponding to the processes $ \mu^+\mu^- \to Zh_0 \to l^{+} l^{-}b \bar{b}$ and $ \mu^+\mu^- \to ZH_0 \to l^{+} l^{-} W^{+} W^{-}$, respectively.
Our results for the main background as well as for the total number of events are shown in Tables~\ref{h00}-\ref{bzH0}. These tables show that the incorporation of the final signal and SM background has an impact on the sensitivity of the processes. Particularly, in the case of the process  $ \mu^+\mu^- \to ZH_0 \to l^{+} l^{-} W^{+} W^{-}$ (see  Table~\ref{bzH0}), a significant improvement in the signal is expected.

Finally, studying the resonances is an excellent place to look for new physics.  In this regard, our results may be helpful to the scientific community and complement other studies performed in extended models. 
Our predictions presented in this work could be relevant for the community to prioritize future searches and experimental efforts.

\vspace{2cm}

\begin{center}
{\bf Acknowledgements}
\end{center}

J. M. Mart\'inez-Mart\'inez is a scholarship fellow of CONAHCyT. E. Cruz-Albaro appreciates the postdoctoral stay at the
Universidad Autónoma de Zacatecas. A.G.R. and M.A.H.R. thank SNII and PROFEXCE (M\'exico).

\vspace{1cm}
%\newpage

\begin{center}
   {\bf Declarations}
\end{center}

 Data Availability Statement: All data generated or analyzed during this study are included in this article.

%\vspace{2cm}

\newpage

\appendix

\section{Feynman rules for the BLHM} \label{rulesF}

In this Appendix, we provide the Feynman rules for the interaction vertices involved in the calculation of the Higgs-strahlung production processes $\mu^+\mu^- \to Zh_0, ZH_0$.

\begin{table}[H]
\caption{Three-point couplings of one gauge boson to two leptons in the BLHM.
\label{boson-fermion}}
\begin{tabular}{|p{2.2cm} p{1.5cm} p{12.2cm}|}
\hline
\hline
\textbf{Particle} &   & \hspace{3.5cm} \textbf{Couplings} \\
\hline
\hline
$Z \bar{e}_i e_i $  & $g_{A}^{Z \bar{e}_i e_i } =$  & $ \frac{i g }{4 c_{W}}  $ \\
$                 $  & $g_{V}^{Z \bar{e}_i e_i  } =$  & $  \frac{i g }{4 c_{W}} (-1+ 4 s^{2}_{W}) $ \\
\hline
\hline
$Z' \bar{e}_i e_i $  & $g_{A}^{Z' \bar{e}_i e_i } =$  & $  \frac{i g c_{g} }{8 s_{g}} \left(2+ \frac{(c_{g}-s_{g}) s_{g}^2
   (c_{g}+s_{g}) \left(c_{W}^2-3 s_{W}^2\right)
   v^2}{c_{W}^2 \left(f^2+F^2\right)}\right)  $ \\
$                 $  & $g_{V}^{Z' \bar{e}_i e_i  } =$  & $  -\frac{i g c_{g}  }{8 s_{g}} \left( 2 + \frac{(c_{g}-s_{g}) s_{g}^2
   (c_{g}+s_{g}) \left(c_{W}^2+s_{W}^2\right)
   v^2}{c_{W}^2 \left(f^2+F^2\right)}\right) $ \\
\hline
\hline
\end{tabular}
\end{table}

\begin{table}[H]
\caption{Three-point couplings of two gauge bosons to one Higgs boson in the BLHM.
\label{2boson-Higgs}}
\begin{tabular}{|p{1.9cm} p{1.5cm} p{12.8cm}|}
\hline
\hline
\textbf{Particle} &   & \hspace{3.5cm} \textbf{Couplings} \\
\hline
\hline
$Z Z h_{0} $  & $g_{Z Z h_{0}} =$  &  $\frac{g m_{W}
   \sin (\alpha +\beta )}{c^{2}_{W}}    -\frac{s_{W}^2 v^3
   \left(g^2+g_{Y}^2\right)^2 \sin (\alpha +\beta )}{6
   g_{Y}^2 f^2} \hfill \break -\frac{s_{W} v^3 x_{s} \left(g^2+g_{Y}^2\right) \sin (\alpha
   +\beta ) \left( - c_{g}^2 g g_{Y}+c_{g} s_{g}
   s_{W} \left(g^2+g_{Y}^2\right)+g g_{Y}
   s_{g}^2\right)}{2 c_{g} s_{g} g_{Y}^2
   \left(f^2+F^2\right)}  $ \\
\hline
\hline
$Z Z' h_{0} $ & $g_{Z Z' h_{0}} =$  &   $ -\frac{g
   s_{W} v \left(c_{g}^2-s_{g}^2\right)
   \left(g^2+g_{Y}^2\right) \sin (\alpha +\beta )}{2 c_{g}
    s_{g} g_{Y} }+  \frac{g s_{W} v^3
   \left(c_{g}^2-s_{g}^2\right) \left(g^2+g_{Y}^2\right)
   \sin (\alpha +\beta )}{6 c_{g} s_{g} g_{Y} f^2 }  \hfill \break  + \frac{v^3
   x_{s} \sin (\alpha +\beta ) \left(c_{g}^2 g g_{Y}
   s_{W} \left(g^2+g_{Y}^2\right)+2 c_{g} s_{g}
   \left(g^4 s_{W}^2+g^2 g_{Y}^2 \left(2
   s_{W}^2+1\right)+g_{Y}^4 s_{W}^2\right)-g g_{Y}
   s_{g}^2 s_{W} \left(g^2+g_{Y}^2\right)\right)}{2
   c_{g}  s_{g}g_{Y}^2 \left(f^2+F^2\right)} $ \\
\hline
\hline
$Z Z H_{0} $  & $g_{Z Z H_{0}} =$  & $    \frac{s_{W}^2 v
   \left(g^2+g_{Y}^2\right)^2 \cos (\alpha +\beta )}{2 g_{Y}^2}  - \frac{s_{W}^2 v^3
   \left(g^2+g_{Y}^2\right)^2 \cos (\alpha +\beta )}{6
   g_{Y}^2 f^2 } \hfill \break -\frac{s_{W} v^3 x_{s} \left(g^2+g_{Y}^2\right) \cos
   (\alpha +\beta ) \left(c_{g}^2 (-g) g_{Y}+c_{g}
   s_{g} s_{W} \left(g^2+g_{Y}^2\right)+g g_{Y}
   s_{g}^2\right)}{2 c_{g}  s_{g}g_{Y}^2
   \left(f^2+F^2\right)}  $ \\
\hline
\hline
$ Z Z' H_{0} $  &  $g_{Z Z' H_{0}} =$  & $ -\frac{g
   s_{W} v \left(c_{g}^2-s_{g}^2\right)
   \left(g^2+g_{Y}^2\right) \cos (\alpha +\beta )}{2 c_{g}
    s_{g} g_{Y} }+\frac{g s_{W} v^3
   \left(c_{g}^2-s_{g}^2\right) \left(g^2+g_{Y}^2\right)
   \cos (\alpha +\beta )}{6 c_{g} s_{g} g_{Y} f^2  } \hfill \break + \frac{v^3
   x_{s} \cos (\alpha +\beta ) \left(c_{g}^2 g g_{Y}
   s_{W} \left(g^2+g_{Y}^2\right)+2 c_{g} s_{g}
   \left(g^4 s_{W}^2+g^2 g_{Y}^2 \left(2
   s_{W}^2+1\right)+g_{Y}^4 s_{W}^2\right)-g g_{Y}
   s_{g}^2 s_{W} \left(g^2+g_{Y}^2\right)\right)}{2
   c_{g}  s_{g} g_{Y}^2 \left(f^2+F^2\right)} $ \\
\hline
\hline
\end{tabular}
\end{table}

%\vspace{2cm}

\newpage

%\begin{figure}
%        \centering
%        \includegraphics[width=0.7\textwidth]{Fig-14.eps}
%        \caption{Correlation between $F$ and $f$. The contours are for $\sigma_{Tot}=0.0018, 0.0014, 0.0010, 0.0006\hspace{0.8mm}\rm fb$.}
%        \label{fig:SeS}
%\end{figure}

\end{document}